\documentclass[aps,superscriptaddress,prc,twocolumn,nofootinbib]{revtex4}

\usepackage{graphicx}
\usepackage{epstopdf}
\usepackage{subfigure}
\usepackage{epsfig}
\usepackage{amsmath,amssymb,amsfonts}
\usepackage{color}
\usepackage[utf8]{inputenc}
\allowdisplaybreaks
\usepackage[bookmarks,
                bookmarksopen = true,
                bookmarksnumbered = true,
                linktocpage,
                colorlinks = true,
                linkcolor = blue,
                urlcolor  = blue,
                citecolor = blue,
                anchorcolor = green,
                hyperindex = true,
                hyperfigures]
                {hyperref}
\usepackage{multirow}
\usepackage{array}
\usepackage{epstopdf}
\usepackage{upgreek}

\begin{document}

\title{Nuclear modification of $B_c$ mesons in relativistic heavy-ion collisions based on a linear Boltzmann transport model}

\author{Lejing Zhang}
\affiliation{Institute of Frontier and Interdisciplinary Science, Shandong University, Qingdao, Shandong 266237, China}

\author{Xiaowen Li}
\affiliation{Institute of Frontier and Interdisciplinary Science, Shandong University, Qingdao, Shandong 266237, China}

\author{Wen-Jing Xing}
\email{wenjing.xing@sdu.edu.cn}
\affiliation{Institute of Frontier and Interdisciplinary Science, Shandong University, Qingdao, Shandong 266237, China}

\author{Shanshan Cao}
\email{shanshan.cao@sdu.edu.cn}
\affiliation{Institute of Frontier and Interdisciplinary Science, Shandong University, Qingdao, Shandong 266237, China}

\author{Guang-You Qin}
\email{guangyou.qin@mail.ccnu.edu.cn}
\affiliation{Institute of Particle Physics and Key Laboratory of Quark and Lepton Physics (MOE), Central China Normal University, Wuhan, 430079, China}

\date{\today}


\begin{abstract}
The nuclear modification factor ($R_\mathrm{AA}$) of $B_c$ mesons in high-energy nuclear collisions provides a novel probe of heavy quark interactions with the quark-gluon plasma (QGP). Based on a linear Boltzmann transport model that incorporates both Yukawa and string types of interactions between heavy quarks and the QGP, we study the production and evolution of heavy quarks and $B_c$ mesons within the same framework. A $B_c$ bound state dissociates while one of its constituent heavy quarks scatters with the QGP with momentum transfer greater than its binding energy. The medium-modified charm and bottom quarks can recombine into $B_c$ mesons, and the medium-modified bottom quarks can also fragment to $B_c$ mesons. We find that most primordial $B_c$ mesons generated from the initial hard collisions dissociate inside the QGP. The production of $B_c$ mesons is primarily driven by the recombination mechanism at low transverse momentum and fragmentation at high transverse momentum. The string interaction dominates over the Yukawa interaction in the nuclear modification of $B_c$ mesons. The participant number dependence of the $B_c$ meson $R_\mathrm{AA}$ is determined by the complicated interplay between the heavy quark yield, energy loss, and the QGP volume. We obtain a reasonable description of the $R_\mathrm{AA}$ of $B_c$ mesons in Pb+Pb collisions at $\sqrt{s_\mathrm{NN}}=5.02$~TeV, and provide predictions for Au+Au collisions at $\sqrt{s_\mathrm{NN}}=200$~GeV.
\end{abstract}

\maketitle


\section{Introduction}
\label{sec:Introduction}

The quark-gluon plasma (QGP) created at the Relativistic Heavy-Ion Collider (RHIC) and the Large Hadron Collider (LHC) is a strongly interacting Quantum Chromodynamics (QCD) matter that exhibits properties of a nearly perfect fluid~\cite{Gyulassy:2004zy,Jacobs:2004qv,Muller:2012zq}. Among various probes of properties of the QGP, heavy quarks are excellent candidates~\cite{Andronic:2015wma}. Due to their large masses that exceed the thermal scale of the QGP, open heavy quarks (charm and bottom quarks) mainly originate from the initial hard nucleus-nucleus collisions and then scatter through the QGP with conserved flavors. Their nuclear modification encodes rich information about the thermodynamic properties of the QGP~\cite{Dong:2019byy,Dong:2019unq,Cao:2018ews,Rapp:2018qla,Xu:2018gux,Liu:2023rfi} and the hadronization mechanism in the presence of a color-deconfined QCD medium~\cite{Zhao:2023nrz}. Meanwhile, the sequential melting of bound states of heavy quarks (charmonia and bottomonia) serve as a QGP thermometer, and has been considered one of the smoking gun signatures of formation of the QGP in high-energy nuclear collisions~\cite{Matsui:1986dk,Satz:2005hx,Vogt:1999cu,Wong:2004zr,Rapp:2008tf,CMS:2012gvv,Andronic:2024oxz}.

The QGP effect on heavy quarks can be quantified by the ratio of their spectra in nucleus-nucleus (A+A) collisions to those in proton-proton ($p+p$) collisions, known as the nuclear modification factor ($R_\mathrm{AA}$). It is now generally accepted that the $R_\mathrm{AA}$ of charmonia results from the interplay between their dissociation and regeneration processes inside the QGP~\cite{Braun-Munzinger:2000csl,Thews:2000rj,Zhao:2011cv,Yao:2018sgn,Chen:2018kfo,Wu:2024gil}. The former causes a suppression of charmonia, while the latter causes an enhancement, both becoming stronger as one moves from peripheral to central heavy-ion collisions, and from the RHIC to LHC environments. Compared to Au+Au collisions at RHIC, the significantly larger yield of charm quark pairs in Pb+Pb collisions at the LHC leads to a much higher charmonium production through regeneration. As a result, one observes a larger $R_\mathrm{AA}$ of $J/\psi$ at the LHC~\cite{ALICE:2013osk} than at RHIC~\cite{PHENIX:2006gsi}; and the $R_\mathrm{AA}$ decreases as the participant number ($N_\text{part}$) increases at RHIC, while shows a weak dependence on $N_\text{part}$ at the LHC. Contribution from regeneration to bottomonium production is much weaker than that to charmonium production~\cite{Du:2017qkv,Yao:2018sgn,Liu:2010ej,CMS:2012gvv} due to the much smaller yield of bottom quarks than charm quarks, although it has been suggested that the correlated recombination process may also be important to bottomonium production~\cite{Yao:2020xzw}.

Between charmonium and bottomonium, there exists a special category of bound states of heavy quarks, $B_c$ mesons, each consisting of one charm ($c$) quark and one anti-bottom ($\bar{b}$) quark (or one $\bar{c}$ quark and one $b$ quark). The $B_c$ mesons are expected to have sizes and binding energies lying between those of $J/\Psi$ and $\Upsilon$~\cite{Eichten:1994gt,Ebert:2002pp,Godfrey:2004ya,Eichten:2019gig}, and therefore providing a valuable supplement to the heavy quarkonium physics~\cite{Schroedter:2000ek,Liu:2012tn}. Since the observation of $B_c$ mesons in the CDF experiment~\cite{CDF:1998axz}, they have generated tremendous interest in heavy quark studies. Recently, the CMS experiment has measured the production and nuclear modification of $B_\mathrm{c}$ mesons in relativistic heavy-ion collisions for the first time~\cite{CMS:2022sxl}, offering a new opportunity to study the evolution of both heavy quark bound states and open heavy quarks inside the QGP. Interestingly, instead of a suppression, an obvious enhancement of $B_c$ mesons ($R_\mathrm{AA}>1$) has been seen in the CMS data at transverse momentum ($p_\mathrm{T}$) below 10~GeV, which also shows a weak dependence on $N_\text{part}$. This suggests that the regeneration of $B_c$ mesons dominates over their dissociation inside the QGP, sparking a series of theoretical efforts to understand the observed phenomena. For instance, the instantaneous coalescence model was developed for studying the regeneration of $B_c$ mesons in Ref.~\cite{Chen:2021uar}. Transport models that include both dissociation and regeneration processes have been developed in Refs.~\cite{Zhao:2022auq,Wu:2023djn} and the effect of non-thermalized bottom quarks on the enhancement of $B_c$ mesons has been explored in Ref.~\cite{Zhao:2022auq}. The statistical hadronization model has also been employed to study the $B_c$ production in Ref.~\cite{Zhao:2024jkz}, which is supplemented by the fragmentation process of $b$ quarks at high $p_\mathrm{T}$.

In this work, we extend the linear Boltzmann transport (LBT) model~\cite{Cao:2016gvr,Luo:2023nsi} previously developed for studying open heavy quarks and high energy jets to bound states of heavy quarks. Both perturbative (Yukawa) and non-perturbative (string) interactions between heavy quarks and the QGP have been included~\cite{Xing:2024qcr}, which provide both the dissociation rate of $B_c$ mesons inside the QGP within the quasifree dissociation picture~\cite{Grandchamp:2001pf} and reliable medium-modified spectra of charm and bottom quarks for generating $B_c$ mesons from these heavy quarks. The instantaneous coalescence model previously developed for producing single-heavy-quark hadrons~\cite{Fries:2008hs,Greco:2003vf,Plumari:2017ntm,Cao:2019iqs} is extended to the formation of $B_c$ mesons from $c$ ($\bar{c}$) and $\bar{b}$ ($b$) quarks, and the fragmentation process from $b$ ($\bar{b}$) quarks to $B_c$ mesons~\cite{Braaten:1993jn,Berezhnoy:1997fp} is also included. With this setup, we are able to provide a simultaneous description of the nuclear modification of single-heavy-quark mesons and $B_c$ mesons. 

The rest of this paper is organized as follows. In Sec.~\ref{sec:pp_Bc}, we discuss the initial production of open heavy quarks and $B_c$ mesons from hard $p+p$ and A+A collisions. In Sec.~\ref{sec:medium}, we develop our transport model to describe scatterings of open heavy quarks and dissociation and regeneration of $B_c$ mesons within the same framework. Numerical results on the $B_c$ meson $R_\mathrm{AA}$ are shown in Sec.~\ref{sec:RAA}, where its dependences on different production mechanisms of $B_c$ mesons and different interaction dynamics between open heavy quarks and the QGP are investigated. In the end, we provide a summary and outlook in Sec.~\ref{sec:summary}.

\section{Initial production of heavy quarks and $B_c$ mesons}
\label{sec:pp_Bc}

\begin{figure}[tbp!]
  \centering
  \includegraphics[width=0.87\linewidth]{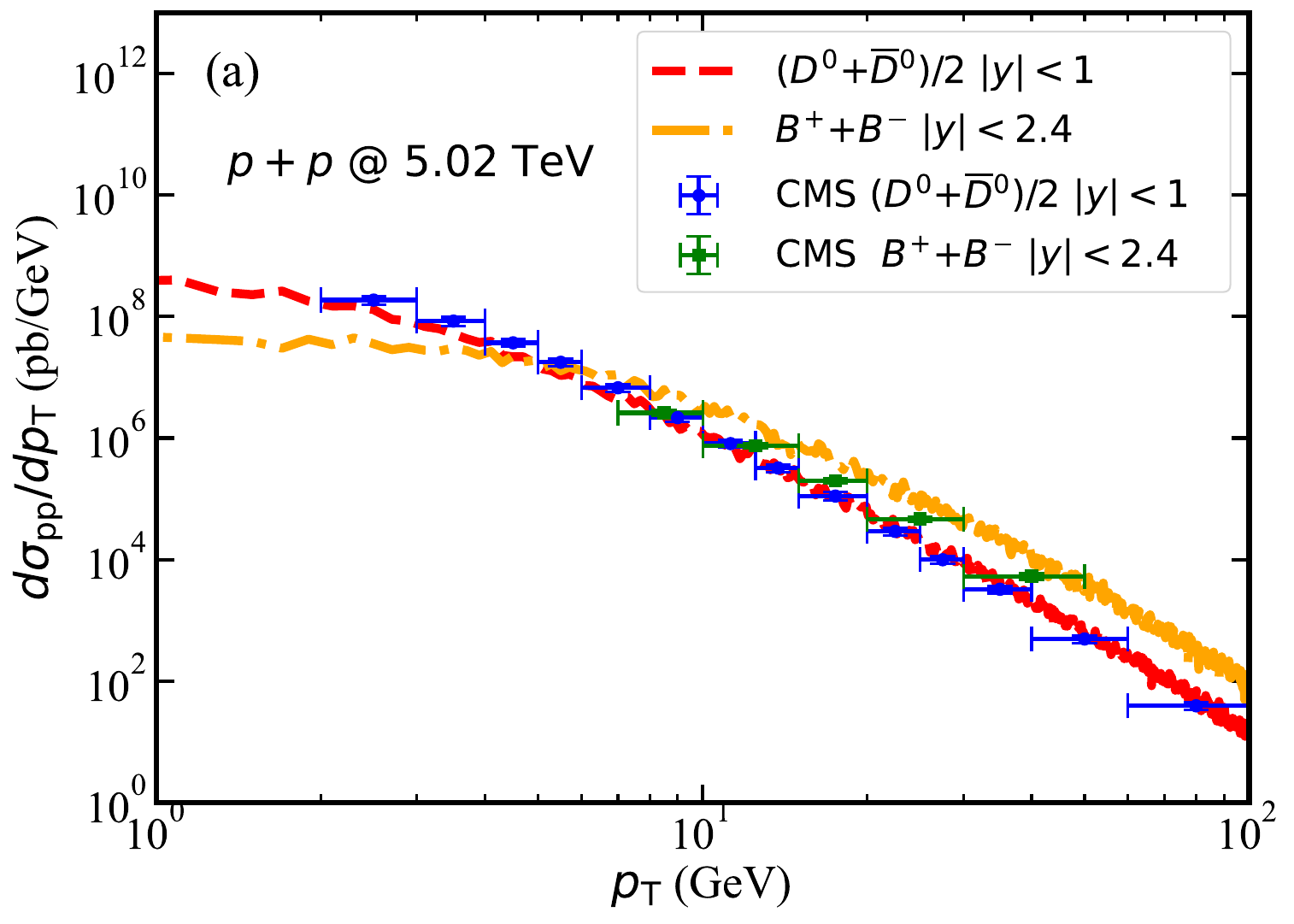}
  \includegraphics[width=0.87\linewidth]{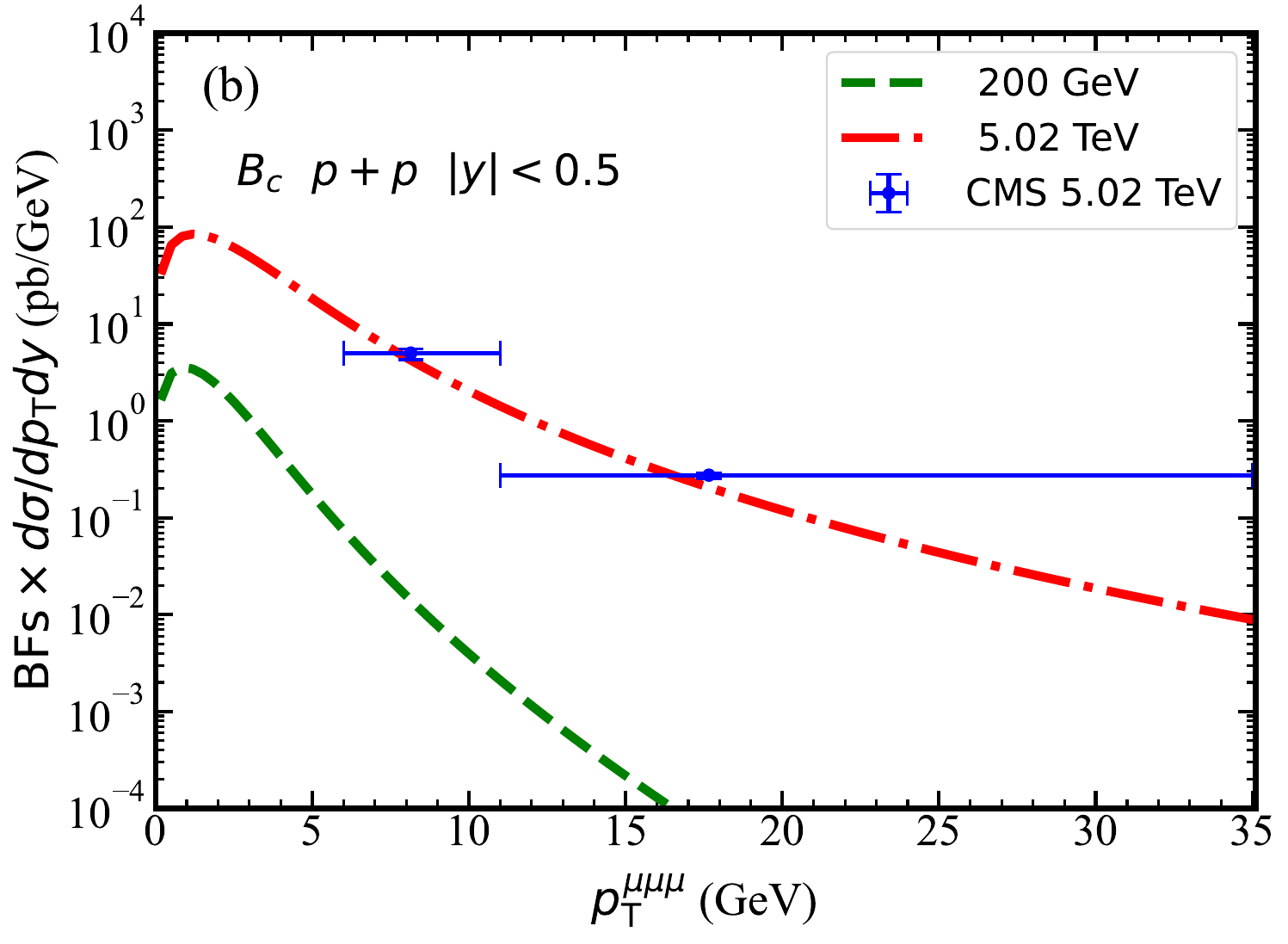}
  \caption{(Color online) $(a)$ The differential cross sections of $D$ and $B$ mesons as functions of $p_\mathrm{T}$ in $p+p$ collisions at $\sqrt{s_\mathrm{NN}}=5.02$~TeV, compared to the CMS data~\cite{CMS:2017qjw,CMS:2017uoy}. $(b)$ The cross sections of $B_c$-decayed $\mu\mu\mu$ as functions of the magnitude of the vector sum of the three muon momenta in $p+p$ collisions at $\sqrt{s_\mathrm{NN}}=5.02$~TeV and 200~GeV, compared to the CMS data at 5.02 TeV~\cite{CMS:2022sxl}.}
  \label{fig:pp}
\end{figure}

We use the fixed-order-next-to-leading log (FONLL) package~\cite{Cacciari:2001td,Cacciari:2012ny,Cacciari:2015fta} to calculate the $p_\mathrm{T}$ spectra (or differential cross sections) of charm and bottom quarks produced from the initial hard nucleon-nucleon scatterings. For $p+p$ collisions, the CT14NLO~\cite{Cacciari:2015fta} parton distribution functions (PDFs) of nucleons are used. By sampling heavy quarks based on their $p_\mathrm{T}$ spectra using the Monte-Carlo (MC) method and then converting them into hadrons using the Pythia (Peterson/SLAC) fragmentation~\cite{Sjostrand:2006za}, one can obtain a reasonable description of the $p_\mathrm{T}$ spectra of $D$ and $B$ mesons in $p+p$ collisions, as shown in Fig.~\ref{fig:pp}(a). For A+A collisions, the cold nuclear matter effects on the $p_\mathrm{T}$ spectra of heavy quarks are introduced by modifying the nucleon PDFs with the EPPS16 parameterizations~\cite{Eskola:2016oht}, and the initial positions of heavy quarks are sampled using the MC-Glauber model~\cite{Miller:2007ri}. These heavy quarks will then interact with the QGP and hadronize via hybrid fragmentation and coalescence models, as will be detailed in the next section.

In $p+p$ collisions, $B_c$ mesons predominantly originate from the fragmentation of bottom quarks, while the contribution from charm quark fragmentation is small~\cite{Braaten:1993jn}. For example, a dominant channel of $B_c$ production is $gg \rightarrow B_c^+ + b + \bar{c}$~\cite{Chang:1992jb,Chang:2003cq,Zhao:2022auq}, where two gluons produce a pair of $b$ and $\bar{b}$ quarks, and the $\bar{b}$ quark emits a gluon that further splits into a pair of $c$ and $\bar{c}$ quarks. The $\bar{b}$ and $c$ quarks combine into a $B_c^+$ meson, leaving a $b$ quark and a $\bar{c}$ quark that can fragment into other hadrons. Similarly, $b$ ($\bar{b}$) quarks generated from the quark annihilation process ($q\bar{q}\rightarrow b\bar{b}$) and the flavor excitation process (e.g., $gb\rightarrow gb$ or $qb\rightarrow qb$) can also produce $B_c^\pm$  mesons in similar ways. 

Effectively, one may construct a fragmentation function for $B_c$ production from $b$ quarks~\cite{Braaten:1993jn}:
\begin{align}
	\label{eq:frag_D}
	D_{b \rightarrow B_c^-}(z) &= N \frac{r z (1-z)^2}{[1-(1-r)z]^6} \nonumber \\
	& \times \left[6 - 18(1-2r)z + (21 - 74r + 68r^2)z^2 \right. \nonumber \\
	& \left. -\, 2(1-r)(6 - 19r + 18r^2)z^3 \right. \nonumber \\
	& \left. +\, 3(1-r)^2(1 - 2r + 2r^2)z^4\right],
\end{align}
and fit it from the experimental data. Here, $z$ represents the $p_\mathrm{T}$ ratio between the $B_c$ meson and its parent $b$ quark, $r=m_c/(m_c+m_{b})$, where the masses of the $c$ and $b$ quarks are set as $m_c=1.27$~GeV and $m_b=4.19$~GeV in this work. The overall constant factor $N=0.002$ is determined from the CMS data of $p+p$ collisions at $\sqrt{s_\mathrm{NN}}=5.02$~TeV~\cite{CMS:2022sxl}, as shown in Fig.~\ref{fig:pp}(b).

Here in Fig.~\ref{fig:pp}(b), we first calculate the $p_\mathrm{T}$ spectrum of $b$ quarks using the FONLL package, and then convolute it with Eq.~(\ref{eq:frag_D}) to obtain the spectrum of $B_c$ mesons:
\begin{eqnarray}
	\label{eq:frag_pT}
\frac{d \sigma_{B c}}{d^{2} p_{\mathrm{T}}^{B_c}}\left(p_\mathrm{T}\right)=\int_{0}^{1} d z \frac{1}{z^{2}} \frac{d \sigma_{b}}{d^{2} p_{\mathrm{T}}^b}\left(\frac{p_\mathrm{T}}{z}\right) D_{b \rightarrow B_c^-}(z).
\end{eqnarray}
The $B_c$ mesons correspond to the sum of $B_c^+$ and $B_c^-$ mesons in both our calculation and the experimental data. Since the experiment measures $B_c$ mesons through the $B_c^+ \to \left( J/\psi \to \mu^- \mu^+ \right) \mu^+ \nu_\mu$ and its charge conjugate channels~\cite{CMS:2022sxl}, the branching factors (BFs) -- $B(B_c^- \to J/\psi \mu^- \bar{\nu}) \approx 1.95\%$~\cite{LHCb:2019tea} and $B(J/\psi \to \mu^+ \mu^-) \approx 5.96\%$~\cite{ParticleDataGroup:2022pth} -- are applied when we compare our cross section of $B_c$ mesons to the data. In addition, a $p_\mathrm{T}$ shift -- $p_\mathrm{T}^{\mu\mu\mu}\approx 0.85\, p_\mathrm{T}^{B_c}$~\cite{CMS:2022sxl} -- has been taken into account. The overall factor $N=0.002$ extracted from this comparison indicates about $0.3\%$ of $b$ quarks fragment to $B_c$ mesons. 

Figure~\ref{fig:pp} suggests the perturbative calculation above can provide a satisfactory description of heavy quark and hadron production in $p+p$ collisions, and therefore establish a reliable baseline for our further study of their nuclear modification in A+A collisions. A prediction of the $B_c$ spectrum in $p+p$ collisions at $\sqrt{s_\mathrm{NN}}=200$~GeV is also provided for future validation. In A+A collisions, the production positions of these initially produced $B_c$ mesons are sampled using the MC-Glauber model before they interact with the QGP.



\section{Medium modification of heavy quarks and $B_c$ mesons}
\label{sec:medium}

\subsection{Linear Boltzmann transport for heavy quark evolution inside the QGP}
\label{sec:model}

We apply the linear Boltzmann transport (LBT) model~\cite{Cao:2016gvr,Luo:2023nsi} to describe the medium modification of charm and bottom quarks inside the QGP. This provides realistic medium-modified charm and bottom quark spectra for studying the regeneration of $B_c$ mesons. The same numerical framework can also be extended to study the dissociation of $B_c$ mesons inside the QGP. In the LBT model, the phase space distribution of heavy quarks $f_a(x_a,p_a)$ evolves according the Boltzmann equation:
\begin{eqnarray}
  \label{eq:boltzmann1}
  p_a \cdot\partial f_a(x_a,p_a)=E_a (\mathcal{C}_\mathrm{el}+\mathcal{C}_\mathrm{inel}).
\end{eqnarray}
Here, $x_a=(t,\vec{x}_a)$ denotes the four-position of heavy quarks and $p_a=(E_a,\vec{p}_a)$ denotes their four-momentum. On the right hand side, the collision integral includes contributions from both elastic and inelastic processes.

For an elastic scattering between a heavy quark $a$ and a thermal parton $b$ inside the QGP to final partons $c$ and $d$, its rate can be extracted from the Boltzmann equation above as 
\begin{align}
  \label{eq:gamma_el}
   \Gamma^\mathrm{el}_{ab \rightarrow cd} & (\vec{p}_a, T) = \frac{\gamma_b}{2E_a}\int \frac{d^3 p_b}{(2\pi)^3 2E_b} \frac{d^3 p_c}{(2\pi)^3 2E_c} \frac{d^3 p_d}{(2\pi)^3 2E_d}\nonumber
  \\ \times\,& f_b (\vec{p}_b, T) \left[1 \pm f_c (\vec{p}_c, T)\right]
   \left[1 \pm f_d (\vec{p}_d, T)\right]\nonumber
  \\ \times\,&\theta \left[s - (m_a + \mu_\mathrm{D})^2\right] \nonumber 
  \\ \times\,& (2\pi)^4 \delta^{(4)}(p_a + p_b - p_c -p_d) \left|\mathcal{M}_{ab \rightarrow cd}\right|^2.
\end{align}
The $\gamma_b$ factor represents the spin-color degeneracy of the medium partons, and their distributions ($f_b$ and $f_d$) take the thermal forms when the local temperature of the medium ($T$) is given. The term $f_c$ can be generally neglected considering the dilute density of heavy quarks inside the QGP compared to that of light partons. The $\pm$ sign takes into account the Bose enhancement and Fermi suppression effects on scatterings, and the $\theta$-function is introduced to require the center-of-mass energy ($\sqrt{s}$) of $a$ and $b$ to be larger than the sum of the heavy quark mass and the Debye screening mass $\mu_\mathrm{D}$. 


The matrix element $\mathcal{M}_{ab \rightarrow cd}$ is evaluated using a potential scattering model, in which an in-medium Cornell-type potential is assumed for interactions between heavy quarks and thermal light partons~\cite{Xing:2021xwc}:
\begin{align}
  V(r)=V_\mathrm{Y}(r)+V_\mathrm{S}(r)=-\frac{4}{3}\alpha_\mathrm{s}\frac{e^{-m_d r}}{r}-\frac{\sigma e^{-m_s r}}{m_{s}},
\label{eq:potential}
\end{align}
where the first term represents the short-range Yukawa interaction, and the second term represents the long-range string interaction. Here, $\alpha_\mathrm{s}$ and $\sigma$ are two parameters that control the coupling strengths associated with the Yukawa and string interactions, respectively; and $m_d$ and $m_s$ are their corresponding screening masses. The screening mass $m_d$ of the Yukawa part is also used as $\mu_\mathrm{D}$ in Eq.~(\ref{eq:gamma_el}). We use the values $\alpha_\mathrm{s}=0.27$, $\sigma=0.45$~GeV$^2$, $m_d=a+bT$, and $m_s=\sqrt{a_s+b_s T}$ with $a=0.20$~GeV, $b=2.0$, $a_s=0$, and $b_s=0.10$~GeV, which provide a reasonable description of the nuclear modification factors and elliptic flow coefficients of $D$ mesons and decay products of $D$ and $B$ mesons at RHIC and LHC~\cite{Xing:2021xwc,Dang:2023tmb,Xing:2024qcr}. The potential obtained using these values also appears consistent with the lattice QCD data~\cite{Xing:2021xwc,Burnier:2014ssa}.

The potential in the momentum space can be obtained from Eq.~(\ref{eq:potential}) via a Fourier transformation as
\begin{eqnarray}
  \label{eq:potential_q}
  V(\vec{q}) = - \frac{4\pi \alpha_\mathrm{s} C_F}{m_d^2+|\vec{q}|^2} - \frac{8\pi \sigma}{(m_s^2+|\vec{q}|^2)^2},
\end{eqnarray}
where $\vec{q}$ represents the momentum exchange between heavy quarks and thermal partons, and $C_F=4/3$ is the color factor. This potential is then employed as an effective gluon propagator between heavy quarks and thermal partons for evaluating their scattering amplitudes:
\begin{align}
  \label{eq:Matrix_cq}
 i \mathcal{M}  =\,& i\mathcal{M_\mathrm{Y}} + i\mathcal{M_\mathrm{S}}\nonumber
 \\=\,& \overline{u}(p') \gamma^{\mu} u(p) V_\mathrm{Y}(\vec{q}) \overline{u}(k') \gamma_{\mu} u(k)\nonumber
 \\&+ \overline{u}(p') u(p) V_\mathrm{S}(\vec{q}) \overline{u}(k') u(k).
\end{align}
Here, we use a vector interaction vertex for the Yukawa part and require it reproduces the perturbative matrix elements of heavy-light-quark scatterings at the leading order~\cite{Combridge:1978kx}, and assume a scalar interaction vertex for the string part~\cite{Riek:2010fk}. As discussed in Ref.~\cite{Xing:2021xwc}, although the form of $V_\mathrm{Y}(r)$ in Eq.~(\ref{eq:potential}) is taken from the quark-anti-quark interaction in the color singlet state of heavy quarkonium~\cite{Megias:2007pq}, an overall factor of $1/(N_c^2-1)$ (with $N_c=3$) is added to $|\mathcal{M}|^2$ for matching its Yukawa part to the perturbative result. Therefore, scattering channels beyond the color singlet channel have been taken into account. Detailed expressions of the scattering matrices can be found in our earlier work~\cite{Xing:2021xwc,Xing:2024qcr}. Note that the Yukawa part of $\left|\mathcal{M}\right|^2$ contains an $\alpha_\mathrm{s}^2$ term. In a $2\rightarrow2$ scattering process, one $\alpha_\mathrm{s}$ is associated with the interaction vertex that connects to the thermal parton inside the QGP, and the other connects to the heavy quark. We take a fixed value $\alpha_\mathrm{s}=0.27$ for the former, while a running value $\alpha_\mathrm{s} = 4\pi / [9 \mathrm{ln}(2ET/\Lambda^2)]$ for the latter, with $\Lambda=0.2$~GeV. We use zero masses for light flavor partons, and 1.27~GeV for charm quarks and 4.19~GeV for bottom quarks.

Using these scattering matrices, we calculate the elastic scattering rate of heavy quarks ($\Gamma^\mathrm{el}_{a,i}$) through a given channel $i$ using Eq.~(\ref{eq:gamma_el}). The total rate is then $\Gamma^\mathrm{el}_a=\sum_i \Gamma^\mathrm{el}_{a,i}$, and the scattering probability during a time step $\Delta t$ is given by $P^\mathrm{el}_a=1-e^{-\Gamma^\mathrm{el}_a\Delta t}$.


For inelastic scatterings, their rate is associated with the number of medium-induced gluons emitted by heavy quarks per unit time as
\begin{eqnarray}
  \label{eq:gamma_inel}
  \Gamma^\mathrm{inel}_a (E_a, T, t) = \int dxdl_{\bot}^2 \frac{dN_g^a}{dxdl_\bot ^2 dt},
\end{eqnarray}
where the medium-induced gluon spectrum is obtained from the higher-twist energy loss calculation~\cite{Zhang:2003wk,Zhang:2018nie,Guo:2000nz,Majumder:2009ge}:
\begin{eqnarray}
  \label{eq:gluon_spectrum}
  \frac{dN_g^a}{dxdl_\bot ^2 dt} = \frac{2C_A \alpha_\mathrm{s} P_a(x) l_{\bot}^4 \hat{q}_a}{\pi (l_{\bot}^2 + x^2 m_a^2)^4} \sin^2 \left(\frac{t-t_i}{2\tau_f}\right).
\end{eqnarray}
Here, $x$ and $l_\perp$ are the energy fraction and the transverse momentum of the emitted gluon relative to the heavy quark, $P_a(x)$ denotes the splitting function of heavy quark, $m_a$ is the heavy quark mass, and the aforementioned running coupling $\alpha_\mathrm{s}$ is used here. The formation time of a medium-induced gluon is given by $\tau_f = 2E_a x(1-x)/(l_{\bot}^2+x^2m_a^2)$, where $t_i$ represents the production time of the heavy quark, or the time when its previous gluon emission happens, and $t$ is the current time. The jet transport coefficient $\hat{q}_a$, which characterizes the transverse momentum broadening square of the heavy quark per unit time due to elastic scatterings, can be obtained using Eq.~(\ref{eq:gamma_el}) with a weight factor $\left[ \vec{p}_c - (\vec{p}_c \cdot \hat{p}_a) \, \hat{p}_a \right]^2$ inserted into the integral. Therefore, effects of the string interaction is included in the medium-induced gluon emission process. With Eq.~(\ref{eq:gamma_inel}), the inelastic scattering probability of a heavy quark within a time step $\Delta t$ is $P^\mathrm{inel}_a=1-e^{-\Gamma^\mathrm{inel}_a\Delta t}$.

\begin{figure}[tbp!]
	\centering
	\includegraphics[width=0.87\linewidth]{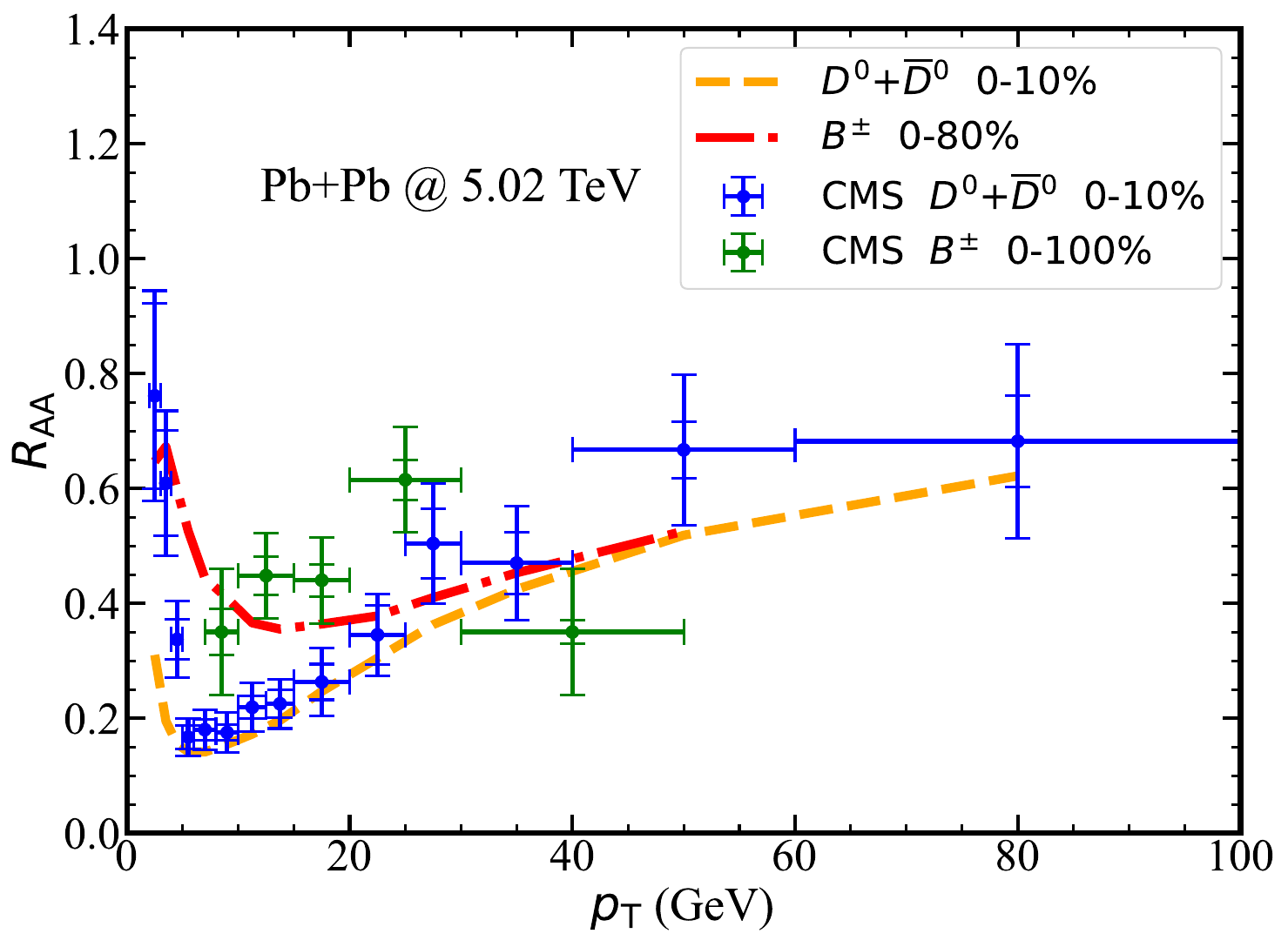}
	\caption{(Color online) The $R_\mathrm{AA}$ of $D$ and $B$ mesons as functions of $p_\mathrm{T}$ in Pb+Pb collisions at $\sqrt{s_\text{NN}}=5.02$~TeV, compared to the CMS data~\cite{CMS:2017qjw,CMS:2017uoy}.}
	\label{fig:AADB}
\end{figure}

To combine elastic and inelastic processes, we add their scattering rates as:
\begin{eqnarray}
	\label{eq:prob_gamma}
	\Gamma^\mathrm{tot}_a=\Gamma^\mathrm{el}_a+\Gamma^\mathrm{inel}_a.
\end{eqnarray}
The total scattering probability then reads:
\begin{eqnarray}
  \label{eq:prob_total}
  P^\mathrm{tot}_a = 1 - e^{-\Gamma_a^\mathrm{tot} \Delta t} = P^\mathrm{el}_a + P^\mathrm{inel}_a - P^\mathrm{el}_a P^\mathrm{inel}_a,
\end{eqnarray}
which can be divided into the probability of pure elastic scattering $P^\mathrm{el}_a(1-P^\mathrm{inel}_a)$ and the probability of inelastic scattering $P^\mathrm{inel}_a$. Using these integrated probabilities and their corresponding differential rates, we use the MC method to simulate the heavy quark evolution through the QGP medium. The medium background, its local temperature and flow velocity information, is generated using the (3+1)-dimensional viscous hydrodynamic model CLVisc~\cite{Pang:2012he,Pang:2018zzo,Wu:2018cpc,Wu:2021fjf}, which has been tuned to describe the soft hadron observables at RHIC and the LHC. Heavy quarks start interacting with the medium at the initial time of its hydrodynamic evolution (0.6~fm), and then scatter though the medium according to our LBT model until the local temperature of the medium drops below their hadronization temperature $T_h$. On the hypersurface of $T_h$, heavy quarks are converted into different species of heavy flavor hadrons. We use the hybrid fragmentation and coalescence model developed in Ref.~\cite{Cao:2019iqs} to produce single-heavy-quark hadrons, e.g., $D$ and $B$ mesons, at $T_h=165$~MeV. This model will also be extended to $B_c$ production later in this work.

As a validation of the LBT model, we present in Fig.~\ref{fig:AADB} the $R_\mathrm{AA}$ of $D$ and $B$ mesons in Pb+Pb collisions at $\sqrt{s_\mathrm{NN}}=5.02$~TeV. With our model setup, a reasonable description of the experimental data can be obtained. This indicates reliable medium-modified spectra of charm and bottom quarks inside the QGP, based on which we can further study their recombination into $B_c$ mesons.

\subsection{Dissociation of $B_c$ mesons}
\label{sec:diss}

The quasifree dissociation method is employed to calculate the dissociation rate of $B_c$ mesons inside the QGP. This method was first developed to study the dissociation of heavy quarkonia~\cite{Grandchamp:2001pf} and then extended to $B_c$ mesons~\cite{Wu:2023djn}. 

We consider the process $i + B_c \rightarrow c + \bar{b} + i \quad (i = q, \bar{q}, g)~$, where the bound state of $B_c$ is destroyed when either of its constituent quarks ($c$ or $\bar{b}$) scatters with a thermal parton inside the QGP ($i$) with a 4-momentum transfer whose magnitude exceeds the binding energy of $B_c$. Therefore, the dissociation rate of $B_c$ mesons reads
\begin{eqnarray}
	\label{eq:P_Bc}
	\Gamma^{\text{diss}}_{B_c}(\vec{p}_{B_c}, T) = \Gamma_c \left( \vec{p}_c, T \right) + \Gamma_{\bar{b}} \left( \vec{p}_{\bar{b}}, T \right),
\end{eqnarray}
in which $\Gamma_c$ and $\Gamma_{\bar{b}}$ can be evaluated using Eq.~(\ref{eq:gamma_el}) with a step function inserted into the integral requiring that the magnitude of the 4-momentum change of the $c$ (or $\bar{b}$) quark is greater than the binding energy of $B_c$ mesons. Since $\vec{p}_{B_c}=\vec{p}_c+\vec{p}_{\bar{b}}=\gamma_c m_c \vec{v}_c + \gamma_{\bar{b}} m_{\bar{b}} \vec{v}_{\bar{b}}$ and $\vec{v}_c=\vec{v}_{\bar{b}}$ inside a $B_c$, we have $\vec{p}_{c}=m_c \vec{p}_{B_c}/(m_c+m_{\bar{b}})$ and $\vec{p}_{\bar{b}}=m_{\bar{b}} \vec{p}_{B_c}/(m_c+m_{\bar{b}})$ in Eq.~(\ref{eq:P_Bc}).

\begin{figure}[tbp!]
  \centering
  \includegraphics[width=0.87\linewidth]{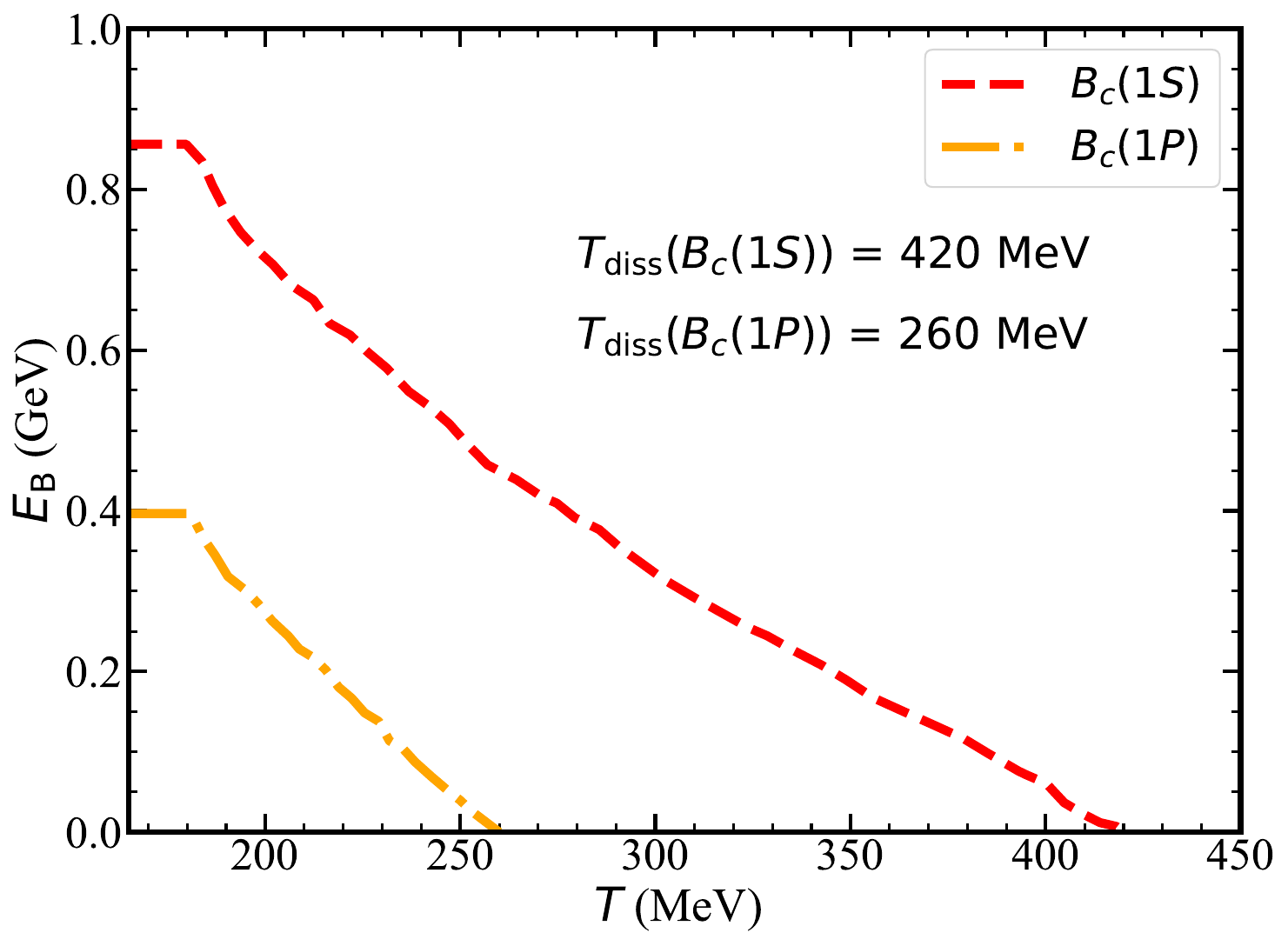}
  \caption{(Color online) The binding energies of the $1S$ and $1P$ states of $B_c$ mesons as functions of the medium temperature, extracted from Ref.~\cite{Wu:2023djn}.}
  \label{fig:Eb}
\end{figure}

In this work, we take the binding energies of $B_c$ mesons from Ref.~\cite{Wu:2023djn}, where the in-medium spectral functions of $B_c$ mesons are evaluated using the $T$-matrix method, and the binding energy of a given $B_c$ state is then given by the difference between the peak position of its spectral function and the sum of the in-medium charm and bottom quark masses. Shown in Fig.~\ref{fig:Eb} are the temperature-dependent binding energies ($E_B$) of the $1S$ and $1P$ states of $B_c$ mesons. In literature, the temperature where $E_B$ becomes zero is called the ``dissociation temperature" of the corresponding state. It is 420~MeV for $B_c(1S)$ and 260~MeV for $B_c(1P)$. Above this value, a $B_c$ meson melts as long as one of its constituent heavy quarks scatters with the medium.


%

\begin{figure}[tbp!]
  \centering
  \includegraphics[width=0.87\linewidth]{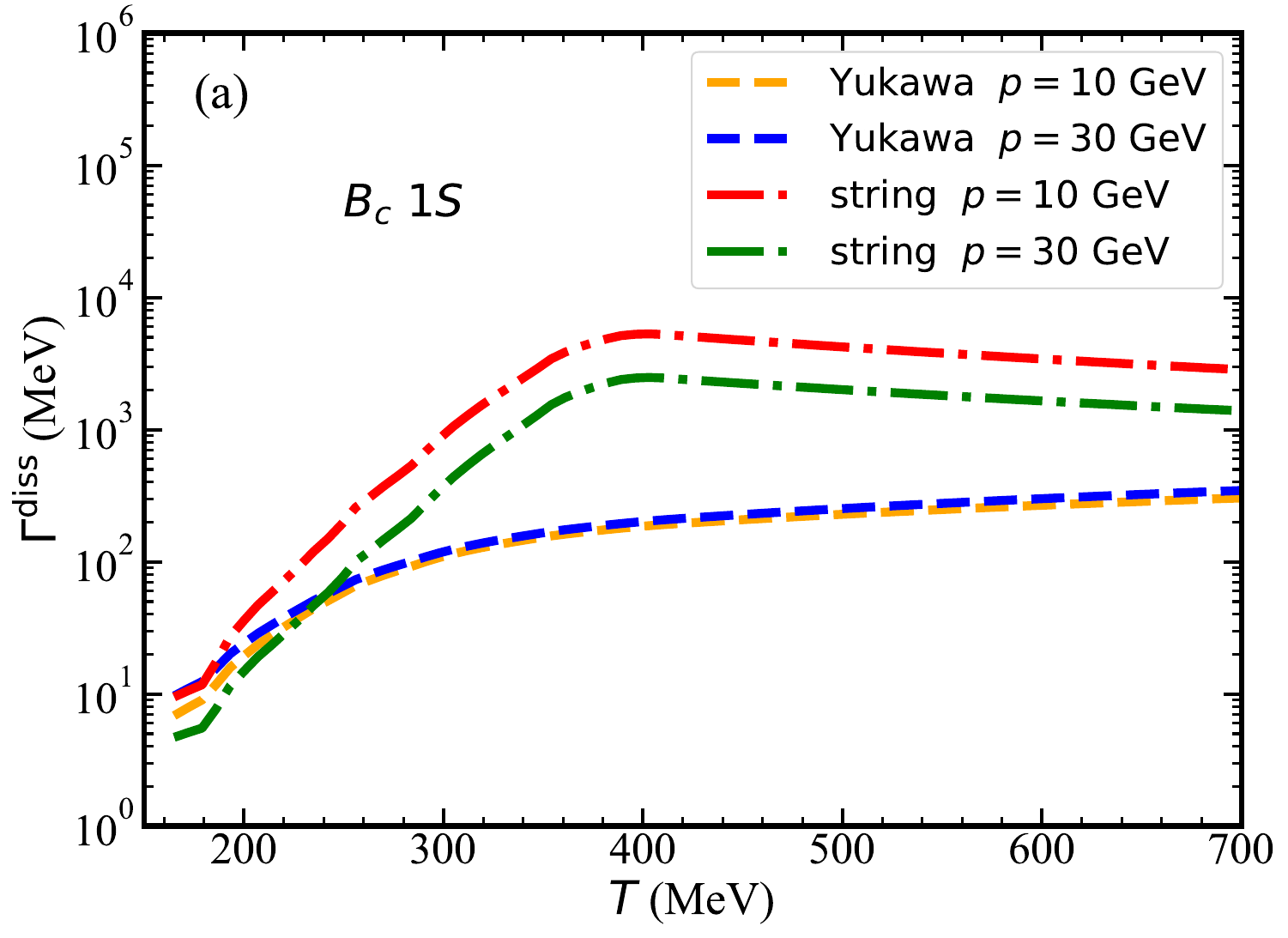}
  \includegraphics[width=0.87\linewidth]{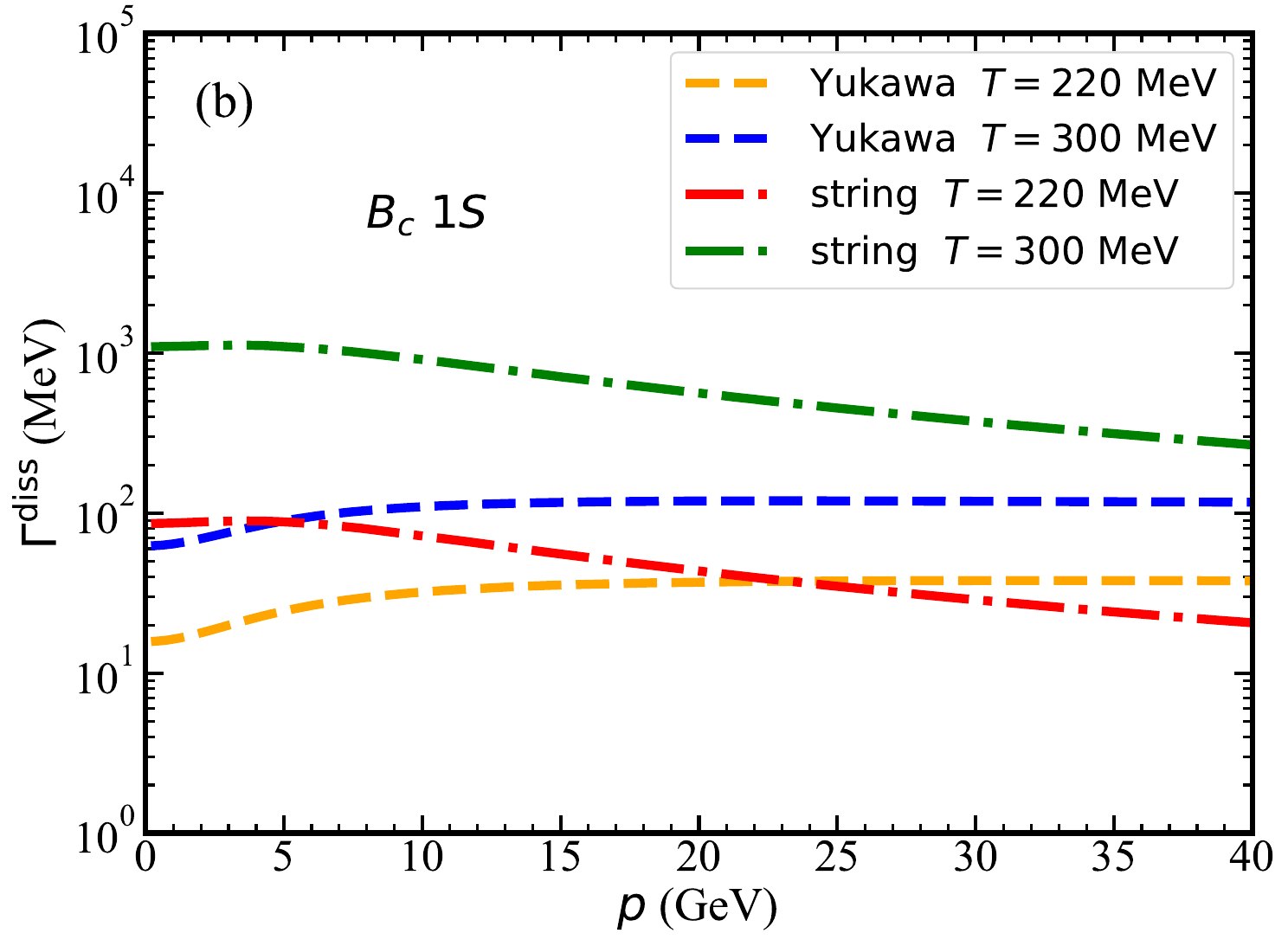}
  \caption{(Color online) The dissociation rates of $B_c(1S)$ mesons as functions of (a) the medium temperature and (b) the $B_c$ momentum, compared between contributions from the Yukawa and string interactions.}
 \label{fig:gamma_bc}
\end{figure}

Using the binding energy, we calculate the dissociation rate of $B_c(1S)$ based on Eq.~(\ref{eq:P_Bc}) and present its dependences on the medium temperature and the momentum of $B_c$ mesons in Fig.~\ref{fig:gamma_bc}. Contributions from the Yukawa and string interactions are compared. In Fig.~\ref{fig:gamma_bc}(a), we see the dissociation rate gradually increases with the medium temperature if only the Yukawa interaction is taken into account, which can be understood with the increasing scattering rate of heavy quarks inside the QGP caused by the increasing density of thermal partons, together with the decreasing binding energy of $B_c$ mesons, as temperature increases. Compared to the Yukawa interaction, the string interaction falls much more quickly as temperature increases, leading to a decrease of the heavy quark scattering rate as temperature increases. This competes with the decreasing binding energy of $B_c$ mesons and results in a non-monotonic temperature dependence of the dissociation rate if only the string interaction is taken into account. Since the string interaction is more sensitive to small momentum exchanges than the Yukawa interaction, we note that the binding energy cut has a stronger impact on the former than the latter when calculating the dissociation rate. In Fig.~\ref{fig:gamma_bc}(b), as the $B_c$ momentum increases, we observe a gradual increase of its dissociation rate due to the Yukawa interaction, while a decrease due to the string interaction, which are in line with the perturbative and non-perturbative features of the Yukawa and string interactions, respectively. In general, the string interaction gives a larger dissociation rate of $B_c$ mesons compared to the Yukawa interaction, especially at low momentum. 


Based on the dissociation rate $\Gamma^\text{diss}$, the dissociation probability for a $B_c$ meson during a time step $\Delta t$ is given by $P^{\text{diss}} = 1 - e^{-\Gamma^\text{diss} \Delta t}$. This is implemented in our LBT model for a MC simulation of $B_c$ mesons inside the QGP. We use the vacuum masses of $B_c$ mesons, 6.324~GeV for their $1S$ state and 6.711~GeV for their $1P$ state predicted by the $T$-matrix calculation~\cite{Wu:2023djn}, to relate their energies and momenta in our simulation.


\subsection{$B_c$ production from medium-modified heavy quarks}
\label{sec:frag_recomb}

Bottom quarks can fragment into $B_c$ mesons after they travel outside the QGP medium, in the same way as they do in $p+p$ collisions. We extract the medium-modified spectrum of bottom quarks at $T_h=165$~MeV, and convert it into the spectra of $B_c$ mesons using Eqs.~(\ref{eq:frag_D}) and~(\ref{eq:frag_pT}).

In relativistic heavy-ion experiments, a large number of charm and bottom quarks can be produced in each collision event, especially at the LHC energy. This enhances the probability of coalescence between a pair of $c$ ($\bar{c}$) and $\bar{b}$ ($b$) quarks into a $B_c^+$ ($B_c^-$) meson. This is also known as the regeneration or recombination process of $B_c$ mesons. To study this process, we extend the instantaneous coalescence model developed for coalescence between a heavy quark and a thermal light quark~\cite{Cao:2019iqs} to coalescence between a $c$ ($\bar{c}$) quark and a $\bar{b}$ (b) quark.

For $B_c$ mesons at a given state, their momentum distribution from recombination can be given by:
\begin{align}
\label{eq:ICM}
\frac{d^{3} N_{B_{c}}}{d^{3} p_{B_c}}=&\,C_\text{rec} \,g_{B_{c}} \int d^{3} {p_{c}} d^{3} {p_{\bar{b}}} \frac{d^{3} N_{c}}{d^{3} {p_{c}}} \frac{d^{3} N_{\bar{b}}}{d^{3} {p_{\bar{b}}}}
\nonumber \\
& \times \delta^{(3)}\left(\vec{p}_{B_c}-\vec{p}_{c}-\vec{p}_{\bar{b}}\right) W(\vec{k}).
\end{align}
Here, $C_\text{rec}$ is an overall normalization factor, which will be extracted from the experimental data later. The statistical factor $g_{B_c}$ takes into account the spin-color degrees of freedom (DOF) of the initial and final states. Since $\text{DOF}=36$ for the initial charm and bottom quarks, and there are 4 $B_c(1S)$ states (1 $^1S_0$ and 3 $^3S_1$) and 12 $B_c(1P)$ states (3 $^1P_1$, 1 $^3P_0$, 3 $^3P_1$, and 5 $^3P_2$), $g_{B_c}=1/9$ for $B_c(1S)$ and 1/3 for $B_c(1P)$. The heavy quark spectra ${d^3N_c}/{d^3 p_c}$ and ${d^3 N_{\bar{b}}}/{d^3p_{\bar{b}}}$ are extracted from our LBT simulation. For simplicity, we take an average hadronization temperature $T_h=220$~MeV for $B_c(1S)$ and $T_h=165$~MeV for $B_c(1P)$, considering that the latter is more loosely bounded than the former. In the end, we assume all $B_c(1P)$ mesons decay into $B_c(1S)$ mesons, and add their spectra together.

\begin{figure}[tbp!]
	\centering
	\includegraphics[width=0.87\linewidth]{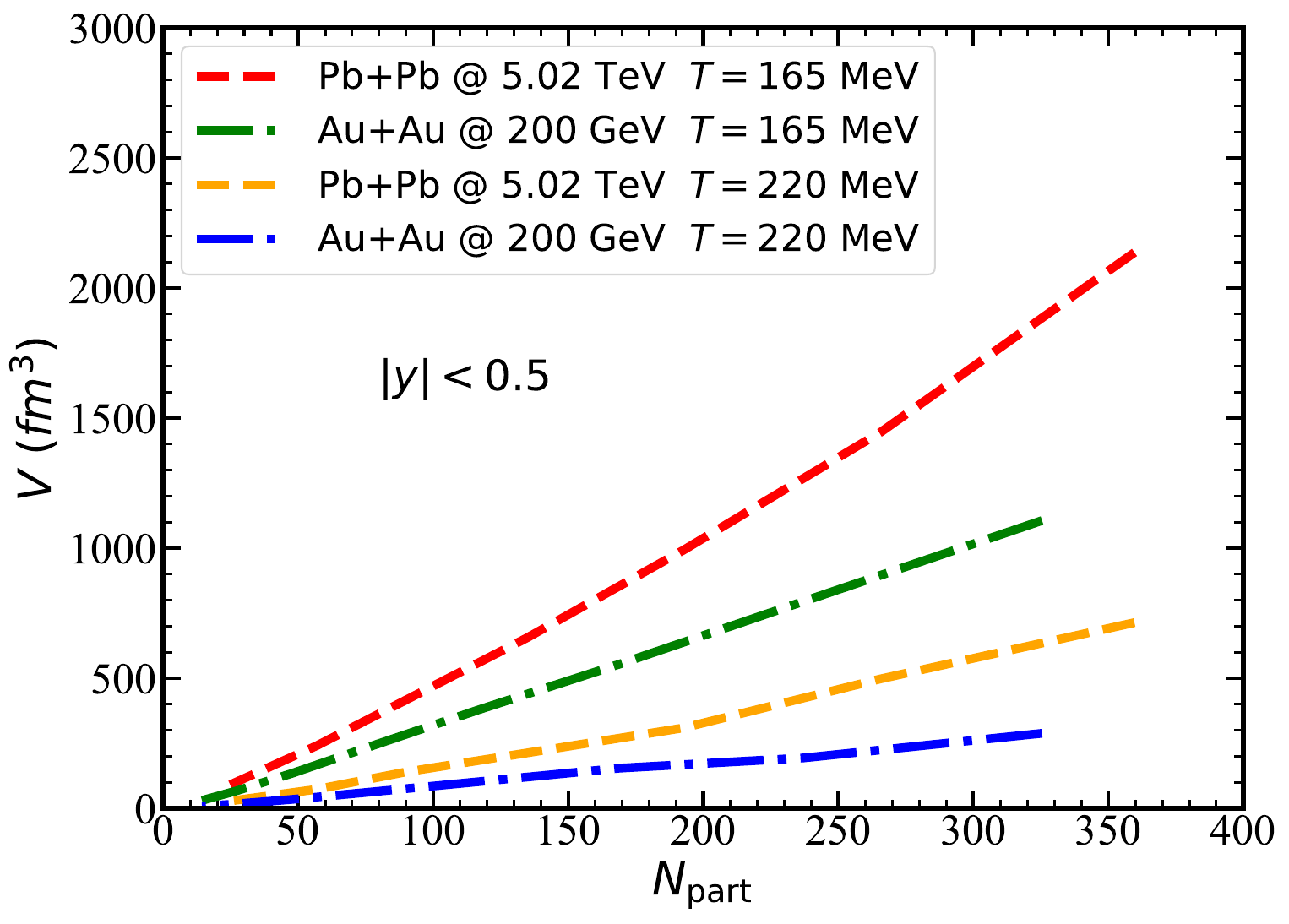}
	\caption{(Color online) The QGP volumes within a spacetime rapidity gap of 1, as functions of the participant number, at different average temperatures in Pb+Pb collisions at $\sqrt{s_\text{NN}}=5.02$~TeV and Au+Au collisions at $\sqrt{s_\text{NN}}=200$~GeV.}
	\label{fig:QGP_Volume}
\end{figure}

In Eq.~(\ref{eq:ICM}), the Wigner function $W(\vec{k})$ provides the probability for a charm quark and an anti-bottom quark to combine into a $B_c$ meson. It is given by the wavefunction overlap between the free quark state and the $B_c$ bound state. By assuming a simple harmonic oscillator potential between the $c$ and $\bar{b}$ quarks, one can obtain the following momentum space Wigner functions for the $1S$ and $1P$ states of a two-body system:
\begin{eqnarray}
\label{eq:winger1}
\mathrm{W}_{s}(\vec{k})&=&\frac{(2 \sqrt{\pi} \sigma)^{3}}{V} e^{-\sigma^{2} k^{2}}, \\
\label{eq:winger2}
\mathrm{~W}_{p}(\vec{k})&=&\frac{(2 \sqrt{\pi} \sigma)^{3}}{V} \frac{2}{3} \sigma^{2} k^{2} e^{-\sigma^{2} k^{2}}.
\end{eqnarray}
Here, we have averaged over the coordinate space with a volume $V$. The relative momentum $\vec{k}$ between the two heavy quarks, $\vec{k}=(m_{\bar{b}} \vec{p}_{c}^{\,\prime}- m_{c} \vec{p}_{\bar{b}}^{\,\prime})/(m_{c} + m_{\bar{b}})$, is defined in their center-of-momentum frame (the rest frame of $B_c$) with $\vec{p}_{c}^{\,\prime}=-\vec{p}_{\bar{b}}^{\,\prime}$. The width parameter $\sigma=1/\sqrt{\mu\omega}$, with $\mu$ the reduced mass of the two-body system and $\omega$ the frequency of the harmonic oscillator, can be related to the mean-square radii of $B_c$ mesons as 
\begin{eqnarray}
	\label{eq:sigma}
	\sigma^{2}(1 S) & = &\frac{4}{3} \frac{\left(m_{c}+m_{\bar{b}}\right)^{2}}{m_{c}^{2}+m_{\bar{b}}^{2}}\left\langle r_{1 S}^{2}\right\rangle, \\
	\sigma^{2}(1 P) & = &\frac{4}{5} \frac{\left(m_{c}+m_{\bar{b}}\right)^{2}}{m_{c}^{2}+m_{\bar{b}}^{2}}\left\langle r_{1 P}^{2}\right\rangle.
\end{eqnarray}
We follow Ref.~\cite{Wu:2023djn} and take $r_{1S}=0.35$~fm and $r_{1P}=0.75$~fm, which are assumed to lie between the radii of charmonium and bottomonium states~\cite{Du:2017qkv}. We have verified that the uncertainties in these radii have minor impact on the $B_c$ spectra obtained from this coalescence model. With Eq.~(\ref{eq:ICM}), the $p_\mathrm{T}$ distribution of $B_c$ mesons is then given by
\begin{eqnarray}
\label{eq:simplify}
\frac{d^{3} N_{B_c}}{d y d^{2}{p_{\text{T}}^{B_c}}}=E_{B_c} \frac{d^{3} N_{B_c}}{d^{3} {p_{B_c}}}.
 \end{eqnarray} 

We note that compared to our earlier work~\cite{Cao:2019iqs} where the coalescence between a heavy quark and a thermal light quark is studied, two differences exist for the coalescence between two heavy quarks here. First, in literature, the coalescence probability between a heavy quark and a thermal light quark is often normalized to 1 for zero  momentum heavy quarks, since flavor is conserved for heavy quarks during their hadronization and low energy heavy quarks cannot fragment. On the other hand, $B_c$ mesons only contribute a tiny fraction to heavy flavor hadrons, and thus one cannot use the normalization condition above to determine the coefficient $C_\text{rec}$ in Eq.~(\ref{eq:ICM}) or $\sigma$ in Eqs.~(\ref{eq:winger1}) and~(\ref{eq:winger2}). In Ref.~\cite{Wu:2023djn}, $C_\text{rec}$ is adjusted according to the thermal limit of the yield of $B_c$ mesons given by the detailed balance between their dissociation and regeneration. However, since the instantaneous coalescence model is not bound by such thermal limit and heavy quarks are not necessarily thermalized with the QGP in realistic heavy-ion collisions, we choose to determine $C_\text{rec}$ from the experimental data on the $B_c$ meson $R_\mathrm{AA}$ later. This is the only parameter we adjust for the coalescence model in this work. Also due to the small number of $B_c$ mesons relative to the total number charm and bottom quarks, we neglect the competition between coalescence and fragmentation processes of producing $B_c$ mesons, and the competition between productions of $B_c$ mesons and other heavy flavor hadrons. 

Second, for coalescence between a heavy quark and a thermal light quark, the volume of the QGP does not affect the yield of heavy flavor hadrons via coalescence, since the proportionality of the light quark spectra $d^3N_l/d^3p_l$ to $V$ in Eq.~(\ref{eq:ICM}) cancels the $V$ in Eq.~(\ref{eq:winger1}) or~(\ref{eq:winger2}). However, the numbers of charm and bottom quarks are determined by the binary collision number in the initial hard nucleus-nucleus collisions, which is not directly proportional to the QGP volume. As a result, the yield of $B_c$ mesons from the coalescence process is sensitive to the QGP volume. This can be reflected in the centrality (or participant number) dependence of the $B_c$ yield, or the difference of its yields between Au+Au and Pb+Pb collisions, as we will discuss in the next section. In Fig.~\ref{fig:QGP_Volume}, we present the volume of the QGP extracted from our hydrodynamic simulations. On the transverse ($z=0$) plane, when the average temperature (converted from the average energy density and the equation of state) of the QGP approaches a given value (e.g. 220~MeV for $B_c(1S)$ production and 165~MeV for $B_c(1P)$ production) at a proper time $\tau$, we extract the area ($A$) of the QGP region and estimate the volume by $V=A\tau\Delta y$, with $\Delta y=1$ the rapidity widow we use to analyze the $B_c$ spectra -- within $y\in(-0.5, 0.5)$. In the figure, we see the QGP volume increases with $N_\text{part}$. It is larger when the average medium temperature is lower, and is larger in Pb+Pb collisions at $\sqrt{s_\text{NN}}=5.02$~TeV than in Au+Au collisions at $\sqrt{s_\text{NN}}=200$~GeV.

\begin{figure}[tbp!]
	\centering
	\includegraphics[width=0.87\linewidth]{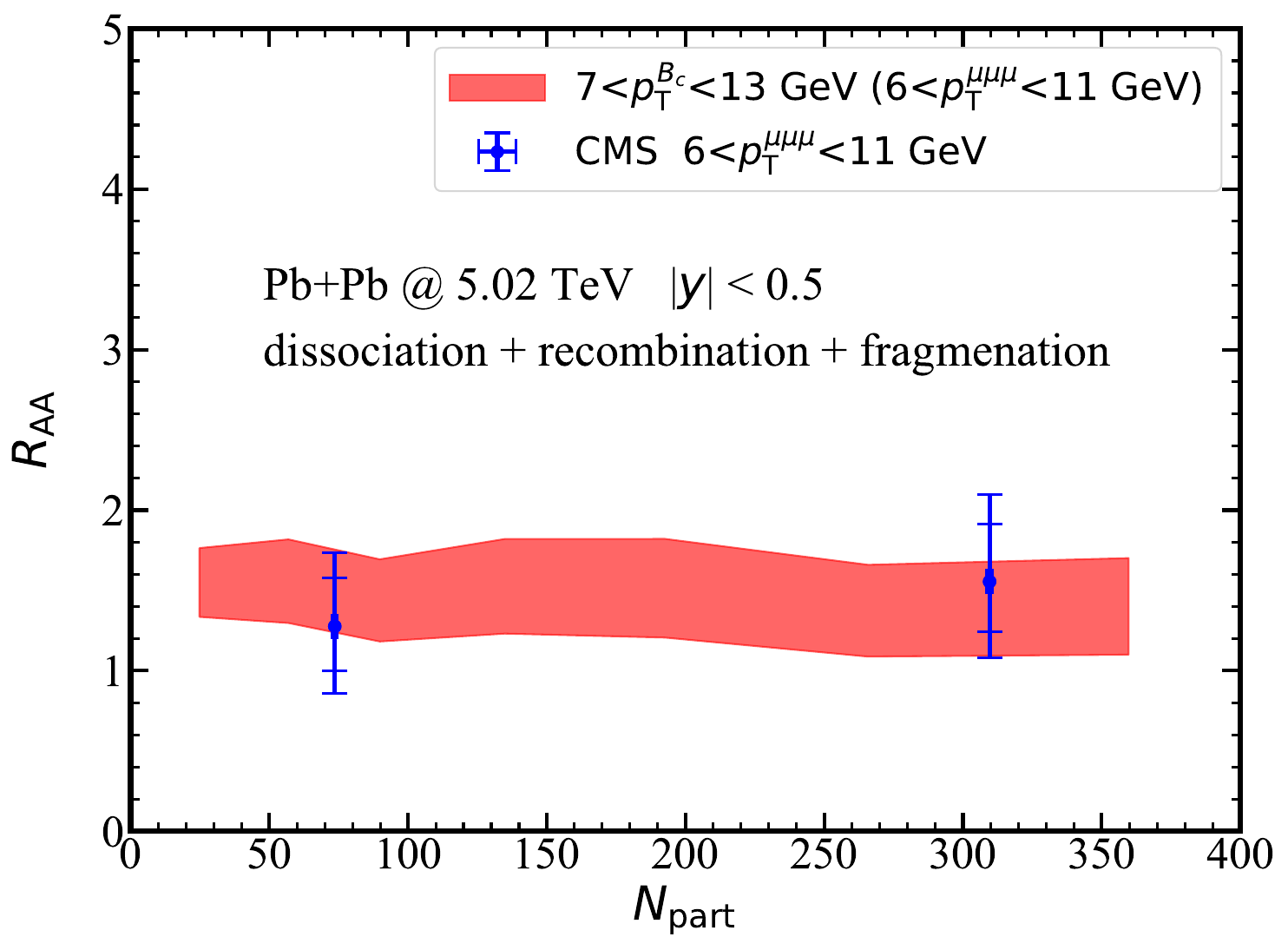}
	\caption{(Color online) The $R_\text{AA}$ of $B_c$ mesons as a function of $N_\text{part}$ in Pb+Pb collisions at $\sqrt{s_\text{NN}}=5.02$~TeV, fit to the CMS data~\cite{CMS:2022sxl}. The band corresponds to a range of $C_\text{rec}\in(3.6, 6.3)$ used in the recombination model.}
	\label{fig:RAA_Npart}
\end{figure}

\section{The nuclear modification factor of $B_c$ mesons}
\label{sec:RAA}

To quantify the QGP effects on $B_c$ mesons in relativistic heavy-ion collisions, we calculate their nuclear modification factor as 
\begin{align}
 R_{\mathrm{AA}}(p_\mathrm{T})\equiv\frac{{d}N^{\mathrm{AA}}/{{d}p_\mathrm{T}}}{{{d}N^{pp}}/{{d}p_\mathrm{T}}\times \left \langle N_\mathrm{coll}\right \rangle},
\end{align}
where $dN^{\mathrm{AA}}/dp_\mathrm{T}$ and $dN^{pp}/dp_\mathrm{T}$ denote their $p_\mathrm{T}$ spectra in A+A and $p+p$ collisions, respectively, and $\langle N_\mathrm{coll} \rangle$ is the average number of binary nucleon-nucleon collisions in each A+A collision. 

In Fig.~\ref{fig:RAA_Npart}, we first present the $p_\mathrm{T}$-integrated $R_\mathrm{AA}$ of $B_c$ mesons as a function of the participant number. In $p+p$ collisions, $B_c$ mesons are produced by $b$ quark fragmentation, as described in Sec.~\ref{sec:pp_Bc}; and in A+A collisions, we include contributions from the primordially produced $B_c$ mesons that survive dissociation, fragmentation of $b$ quarks after medium modification, and recombination between medium-modified $c$ and $b$ quarks, as described in Sec.~\ref{sec:medium}. The model parameter $C_\text{rec}$ in Eq.~(\ref{eq:ICM}) is determined by ensuring that our model's output for the $B_c$ meson $R_\mathrm{AA}$ falls within the error bars of the experimental data~\cite{CMS:2022sxl}. To take into account the 15\% shift of $p_\mathrm{T}$ from a $B_c$ meson to three muons, we integrate the spectra of $B_c$ mesons within $7<p_\mathrm{T}^{B_c}<13$~GeV when comparing their $R_\mathrm{AA}$ to the data analyzed within $6<p_\mathrm{T}^{\mu\mu\mu}<11$~GeV. In the end, $C_\text{rec}$ is constrained within $(3.6, 6.3)$. This range is compatible with the normalization factors reported in Ref.~\cite{Wu:2023djn} if a similar QGP volume is used.


\begin{figure}[tbp!]
	\centering
	\includegraphics[width=0.87\linewidth]{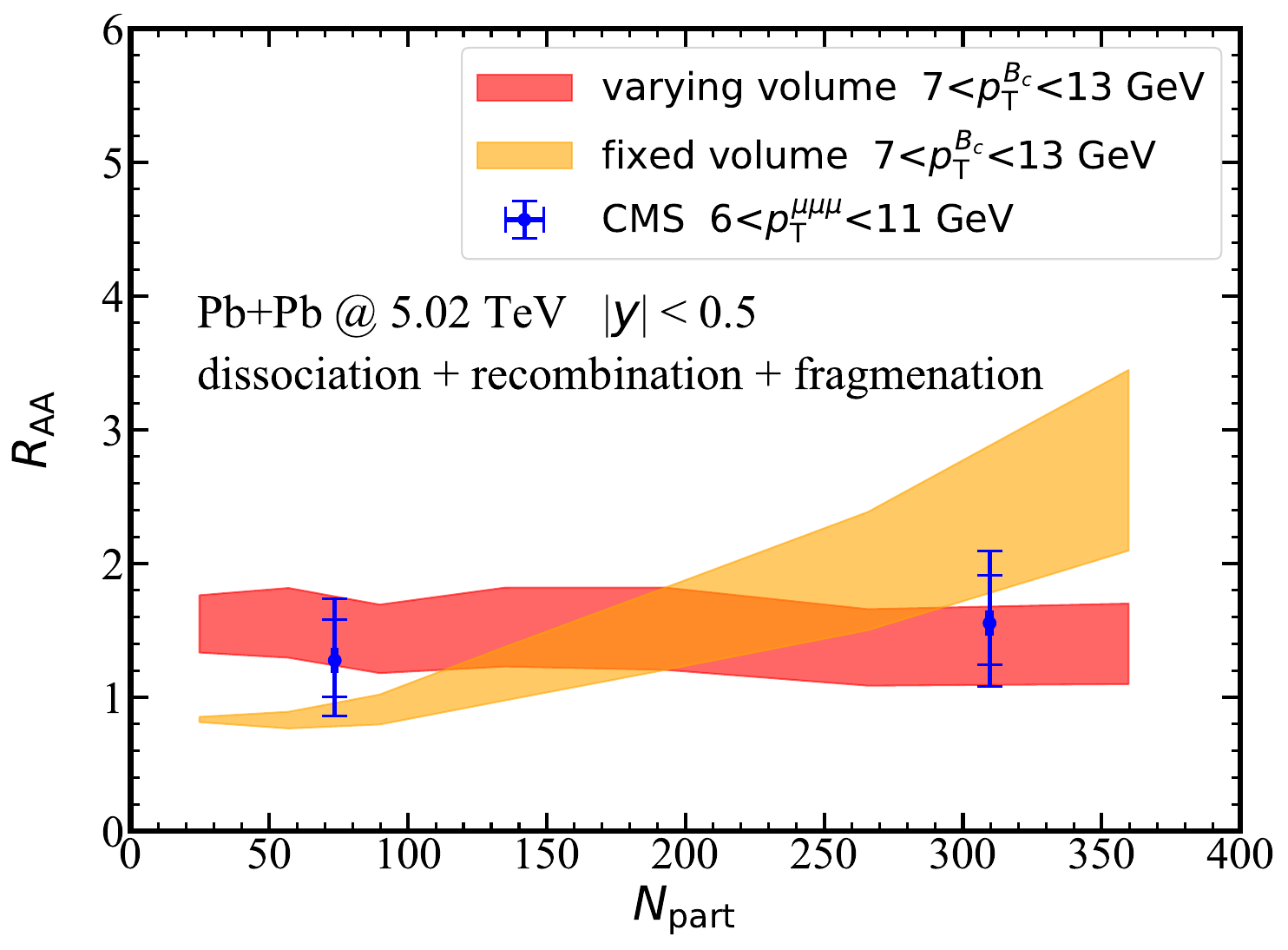}
	\caption{(Color online) The $R_\text{AA}$ of $B_c$ mesons as a function of $N_\text{part}$ in Pb+Pb collisions at $\sqrt{s_\text{NN}}=5.02$~TeV, compared between using varying and fixed volumes of the QGP in the recombination model.}
	\label{fig:RAA1}
\end{figure}

The participant number dependence of the $B_c$ meson $R_\mathrm{AA}$ is affected by several competing factors. As $N_\text{part}$ increases, more heavy quarks are produced in each heavy-ion collision event, leading to an enhanced production of $B_c$ mesons through coalescence. On the other hand, as discussed in Sec.~\ref{sec:frag_recomb}, the QGP volume increases as $N_\text{part}$ increases, which suppresses the coalescence probabilities between charm and bottom quarks, or the Wigner functions. Meanwhile, in more central collisions, one expects stronger dissociation of the primordially produced $B_c$ mesons, and stronger energy loss of charm and bottom quarks that influences the $B_c$ spectrum through coalescence. The combination of these effects lead to a weak dependence of the $B_c$ meson $R_\mathrm{AA}$ on $N_\text{part}$ within the current kinematic range explored at CMS. As discussed in Sec.~\ref{sec:frag_recomb}, the volume dependence is a unique feature of the coalescence production of $B_c$ mesons, which is absent in the production of single-heavy-quark hadrons. In Fig.~\ref{fig:RAA1}, we investigate how the QGP volume affects the $R_\mathrm{AA}$ of $B_c$ mesons. If fixed volumes are used instead of varying volumes with $N_\text{part}$, one would observe an increasing $R_\mathrm{AA}$ of $B_c$ mesons as $N_\text{part}$ increases. Here, for illustrative purposes, we use fixed volumes of 310~fm$^3$ for $B_c(1S)$ and 990~fm$^3$ for $B_c(1P)$, corresponding to the values at $N_\text{part}\approx 193$ (20-30\% centrality) in Fig.~\ref{fig:QGP_Volume}. Such volume dependence of coalescence can be verified with more precise experimental data in the future. 


\begin{figure}[tbp!]
	\centering
	\includegraphics[width=0.87\linewidth]{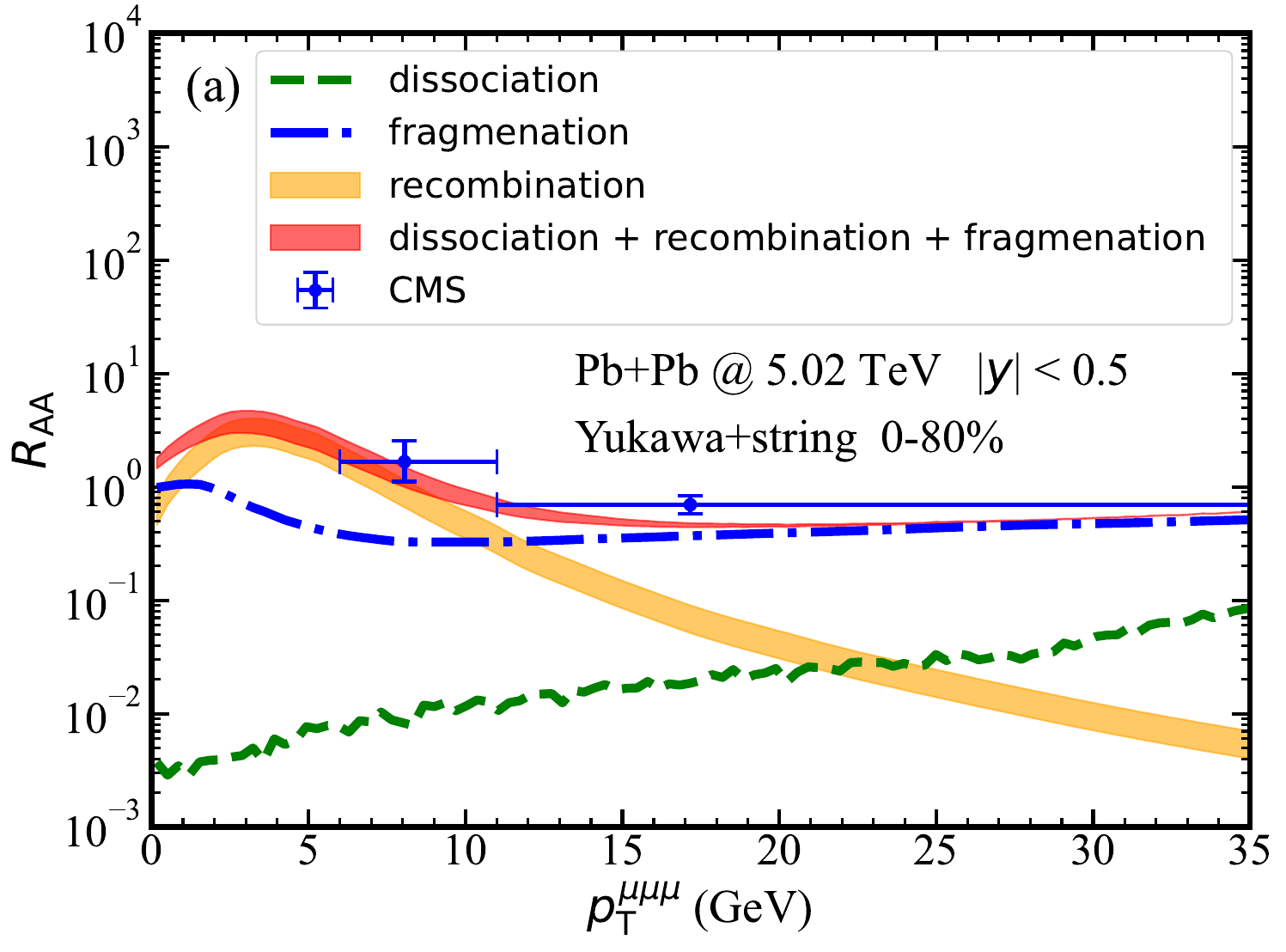}
	\includegraphics[width=0.87\linewidth]{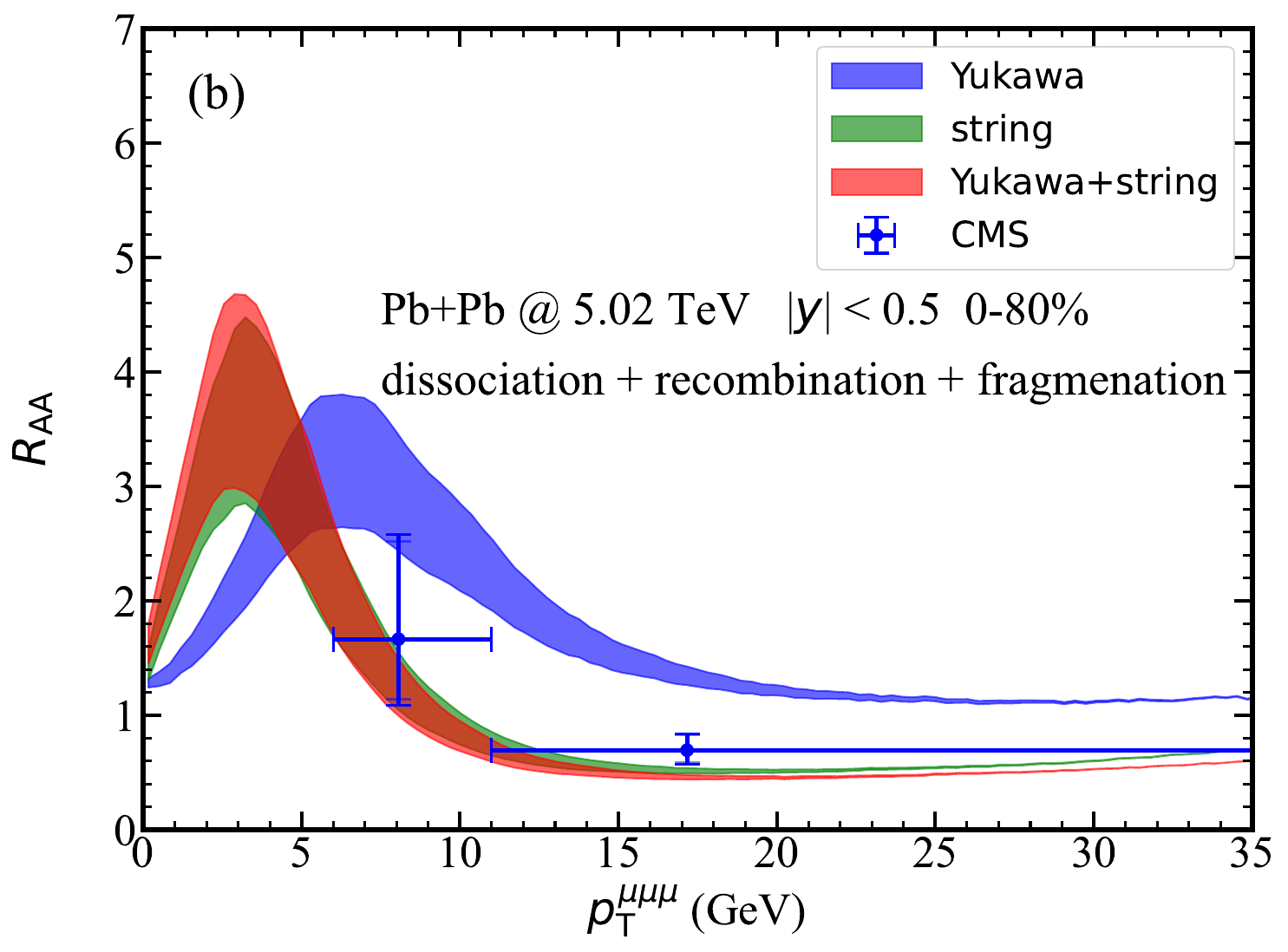}
	\caption{(Color online) The $p_\mathrm{T}$-dependent ${R_\text{AA}}$ of $B_c$ mesons as functions of $p_\text{T}^{\mu\mu\mu}$ in Pb+Pb collisions at $\sqrt{s_\text{NN}}=5.02$~TeV, compared between contributions from (a) the dissociation, fragmentation, and coalescence processes, and (b) the Yukawa and string interactions between heavy quarks and the QGP. The CMS data~\cite{CMS:2022sxl} are shown for comparison.}
	\label{fig:RAA_pT}
\end{figure}

\begin{figure}[tbp!]
	\centering
	\includegraphics[width=0.87\linewidth]{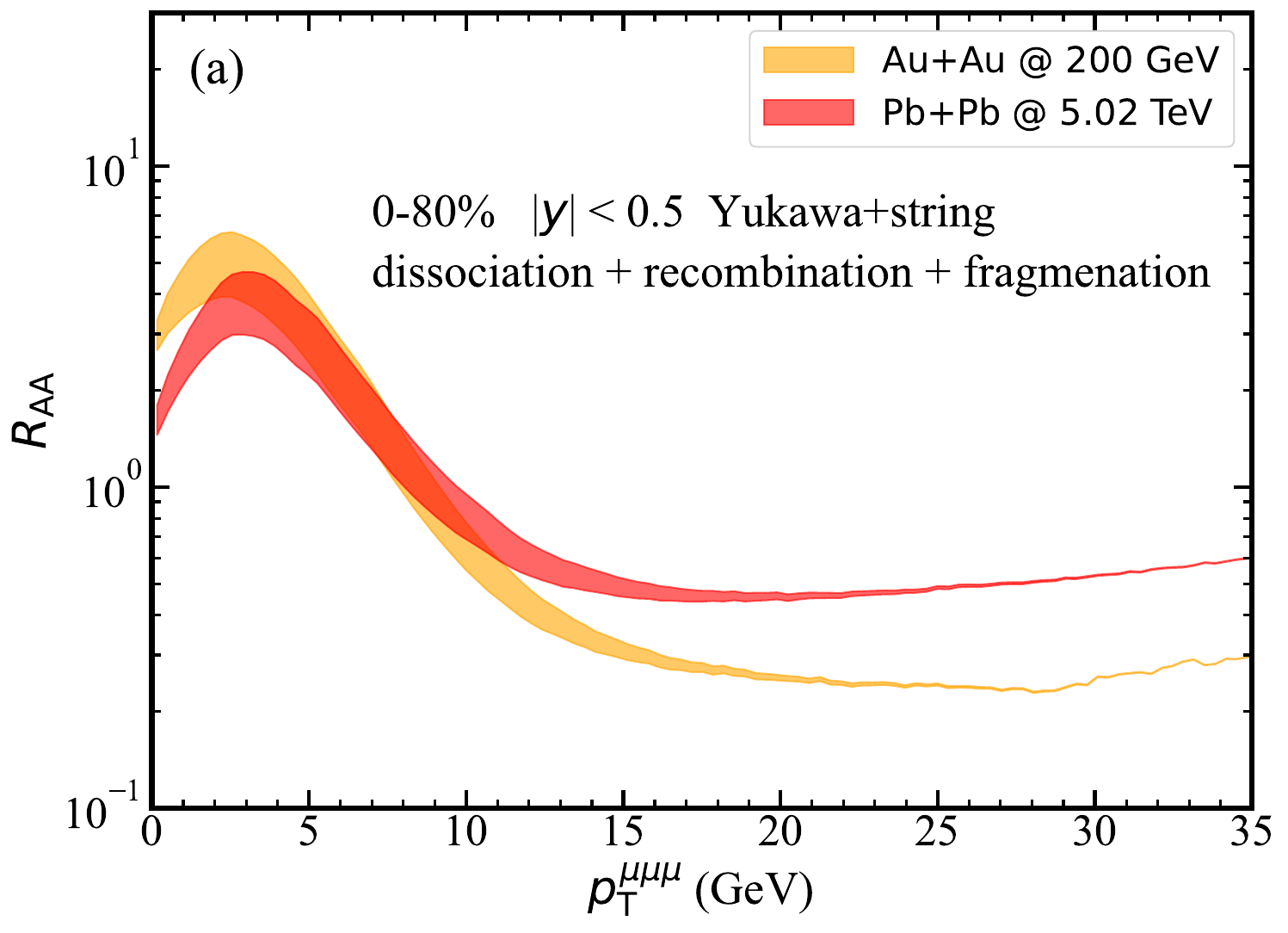}
	\includegraphics[width=0.87\linewidth]{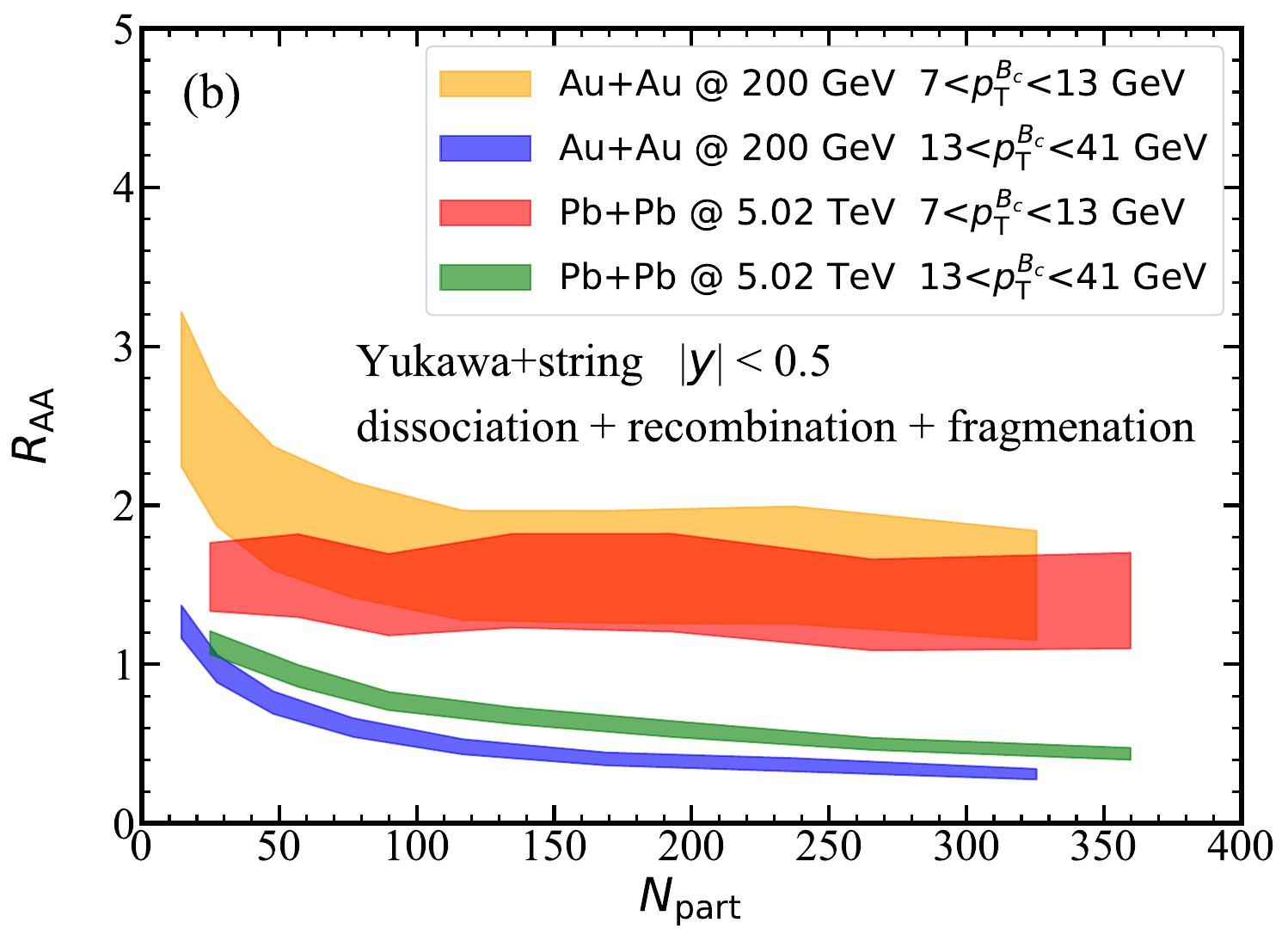}
	\caption{(Color online) Predictions of the $B_c$ meson $R_\text{AA}$ in Au+Au collisions at $\sqrt{s_\text{NN}}=200$~GeV and their comparisons to results in Pb+Pb collisions at $\sqrt{s_\text{NN}}=5.02$~TeV: (a) for their dependences on $p_\mathrm{T}$, and (b) for their dependences on $N_\text{part}$ within two $p_\text{T}$ ranges.}
	\label{fig:200_RAA}
\end{figure}

In Fig.~\ref{fig:RAA_pT}, we further study the $p_\mathrm{T}$-differentiated $R_\mathrm{AA}$ of $B_c$ mesons in Pb+Pb collisions at $\sqrt{s_\text{NN}}=5.02$~TeV, and its dependences on different production mechanisms of $B_c$ mesons in heavy-ion collisions and different interaction dynamics of heavy quarks inside the QGP. In Fig.~\ref{fig:RAA_pT}(a), one can observe that the majority of the initially produced $B_c$ mesons dissociate inside the QGP when both Yukawa and string interactions are included in our model calculation. The final state $B_c$ mesons in heavy-ion collisions are mainly produced by fragmentation of medium-modified bottom quarks and coalescence between medium-modified charm and bottom quarks: the former dominates at high $p_\mathrm{T}$ while the latter dominates at low $p_\mathrm{T}$. The bands here correspond to the range of $C_\text{rec}\in(3.6, 6.3)$ we extracted for the coalescence process from the $p_\mathrm{T}$-integrated $R_\mathrm{AA}$ earlier. We note that our current calculation of dissociation involves certain simplifications. All $B_c$ mesons initially produced from vacuum fragmentation of $b$ quarks are assumed to be in the $1S$ state, dissociation of the $1P$ state has not been included. Besides, there is double counting of $B_c$ mesons produced from fragmentation of both initially generated $b$ quarks and medium-modified $b$ quarks. A more rigorous calculation should consider the formation times of different states of $B_c$ mesons in vacuum, and only take into account dissociation of the $B_c$ mesons that form before the initial time of the QGP medium. Furthermore, inelastic scatterings between heavy quarks and the QGP can also cause dissociation of $B_c$ mesons, which is beyond the $i + B_c \rightarrow c + \bar{b} + i $ process we consider in the present work. However, since only a tiny fraction of $b$ quarks can fragment into $B_c$ mesons and most primordial $B_c$ mesons are destroyed by the QGP, these simplifications have negligible impact on our final results and will be improved in our future efforts.


Contributions from the Yukawa and string interactions between heavy quarks and the QGP to the $B_c$ meson $R_\mathrm{AA}$ are compared in Fig.~\ref{fig:RAA_pT}(b). Compared to the Yukawa interaction, the much more intense string interaction leads to not only stronger dissociation of the primordially produced $B_c$ mesons, but also enhanced energy loss of heavy quarks that results in an increased $c+\bar{b}\rightarrow B_c$ coalescence at low $p_\mathrm{T}$ and a suppressed $\bar{b}\rightarrow B_c$ fragmentation at high $p_\mathrm{T}$. Within our model, the string interaction dominates the $B_c$ meson $R_\mathrm{AA}$.  

In the end, we provide predictions of the $B_c$ meson $R_\mathrm{AA}$ in Au+Au collisions at the RHIC energy, and compare them with the current LHC results. Shown in Fig.~\ref{fig:200_RAA}(a) are the $p_\mathrm{T}$ dependences of the $B_c$ meson $R_\mathrm{AA}$ at the RHIC and LHC energies. At low $p_\mathrm{T}$ where the coalescence mechanism dominates, we observe larger $R_\mathrm{AA}$ in Au+Au collisions at the RHIC energy than in Pb+Pb collisions at the LHC energy. This is opposite to one's expectation that the less abundant heavy quarks produced at the RHIC energy should produce less $B_c$ mesons through coalescence than at the LHC energy. Indeed, we have verified that the yield of $B_c$ mesons through coalescence at RHIC is smaller than that at the LHC. However, their difference is reduced by the smaller QGP volume at RHIC than at the LHC. Moreover, due to the much lower $p+p$ baseline of the $B_c$ spectra at RHIC than at the LHC, as shown in Fig.~\ref{fig:pp}(b), one would obtain a larger $R_\mathrm{AA}$ at RHIC than at the LHC at low $p_\mathrm{T}$. At high $p_\mathrm{T}$ where the $b$ quark energy loss and fragmentation dominates, we observe smaller $R_\mathrm{AA}$ at RHIC than at the LHC. This may also be opposite to one's expectation that $b$ quarks lose less energy and is less quenched at RHIC. However, the much softer spectra of $b$ quarks (or $B_c$ mesons) at RHIC, as shown in Fig.~\ref{fig:pp}(b), can lead to a smaller $R_\mathrm{AA}$ at RHIC than at the LHC even though the energy loss of $b$ quarks has the opposite effect. The same conclusions can be drawn from Fig.~\ref{fig:200_RAA}(b) where the $p_\mathrm{T}$-integrated $R_\mathrm{AA}$'s are shown as functions of $N_\text{part}$ for different $p_\mathrm{T}$ ranges. Within $7<p_\mathrm{T}^{B_c}<13$~GeV where coalescence has a strong impact, the $R_\mathrm{AA}$ at RHIC is larger than that at the LHC. In this region, the $R_\mathrm{AA}$ exhibits a weak dependence on $N_\text{part}$ due to the interplay between the heavy quark abundance and the QGP volume in coalescence, as well as effects of the $B_c$ meson dissociation and the $b$ quark energy loss and fragmentation. Within $13<p_\mathrm{T}^{B_c}<41$~GeV where fragmentation dominates, the $R_\mathrm{AA}$ at RHIC is smaller than that at the LHC. In this region, the $R_\mathrm{AA}$ decreases as $N_\text{part}$ increases because of the stronger $b$ quark energy loss in more central collisions.


\section{Summary and outlook}
\label{sec:summary}

We develop a comprehensive model calculation on the nuclear modification of $B_c$ mesons in relativistic heavy-ion collisions. The spectra of charm and bottom quarks produced from the initial hard collisions are evaluated using a perturbative calculation, and the spectra of $B_c$ mesons in $p+p$ collisions are obtained by convoluting the $b$ quark spectra with a fragmentation function. Heavy quark evolution inside the QGP medium is simulated using the LBT model that includes both elastic and inelastic scatterings processes, in which both Yukawa and string types of interactions are taken into account. This LBT model is extended to describe the melting of the primordially produced $B_c$ mesons based on the quasifree dissociation picture. In the end, $B_c$ mesons are also produced from the medium-modified heavy quarks via both coalescence and fragmentation processes.

We find that the $B_c$ mesons produced by the initial hard collisions have negligible contribution to the final yield of $B_c$ mesons in relativistic heavy-ion collisions due to their large dissociation rate inside the QGP when the string interaction is included. In heavy-ion collisions, the production of $B_c$ mesons is dominated by coalescence between medium-modified charm and bottom quarks at low $p_\mathrm{T}$, while it is dominated by fragmentation of medium-modified bottom quarks at high $p_\mathrm{T}$. The string interaction dominates the nuclear modification of $B_c$ mesons over the Yukawa interaction in our model. The participant number dependence of the $B_c$ meson $R_\text{AA}$ relies on the competition between the heavy quark abundance and the QGP volume in coalescence, together with the dissociation of $B_c$ mesons and the energy loss of heavy quarks. The $R_\text{AA}$ weakly depends on $N_\text{part}$ at low $p_\text{T}$, while decreases as $N_\text{part}$ increases at high $p_\text{T}$. We obtain a reasonable description of the $B_c$ meson $R_\mathrm{AA}$ in Pb+Pb collisions at $\sqrt{s_\text{NN}}=5.02$~TeV, and predict that their $R_\mathrm{AA}$ should be larger at low $p_\mathrm{T}$ while smaller at high $p_\mathrm{T}$ in Au+Au collisions at $\sqrt{s_\text{NN}}=200$~GeV, mainly due to different heavy quark and $B_c$ spectra in $p+p$ collisions at different beam energies. This may be tested by the RHIC experiments in the near future. We note that the cross section of minimum bias inelastic Au+Au collisions at $\sqrt{s_\text{NN}}=200$~GeV is about 7 b, and our current calculation indicates the cross section of $B_c$ production in these minimum bias Au+Au collisions can reach about 30~$\upmu$b. If over 100 billion Au+Au events can be collected by STAR and sPHENIX during their last runs, more than 400k $B_c$ mesons can be produced. Therefore, measurement of $B_c$ mesons at RHIC is possible, especially if the $B_c\rightarrow\mu\mu\mu$ channel can be combined with other decay channels.

While our work achieves a simultaneous description of the nuclear modification of single-heavy-quark mesons and $B_c$ mesons in energetic nuclear collisions, it needs improvement in several directions. Unlike the heavy quark evolution part which is fully implemented using the MC method in the LBT model, only the dissociation process of $B_c$ mesons is realized in MC simulation. For the coalescence process, semi-analytical calculation of the $B_c$ spectra is conducted based on the medium-modified charm and bottom quark spectra extracted from the LBT model. This requires a rough estimation of the average temperature and QGP volume when coalescence happens, which may introduce non-negligible uncertainties to the yield of $B_c$ mesons due to the sensitivity of coalescence to the QGP volume. In addition, the current implementation cannot take into account secondary dissociation of $B_c$ mesons that are regenerated inside the QGP, which could be important. A full MC simulation where heavy quarks and $B_c$ mesons can continue converting to each other till the end of the QGP phase is desired. Furthermore, extension of the $B_c$ production and evolution beyond their $1S$ and $1P$ states may also be necessary~\cite{Chang:2015qea,Zhao:2022auq}. These will be implemented in our follow-up study.

\section*{Acknowledgments}

We are grateful for the generous help of Biaogang Wu, Ralf Rapp, and Jiaxing Zhao in developing our model, and the valuable discussions with Rui-Qin Wang, Qian Yang, Long Ma, Ze-Fang Jiang, Min He, and Xin-Nian Wang. This work is supported by the National Natural Science Foundation of China (NSFC) under Grant Nos.~12175122, 2021-867, 12225503, 11935007, and~12321005. W.-J. Xing  is also supported by the China Postdoctoral Science Foundation under Grant No.~2023M742099.

\bibliographystyle{h-physrev5}
\bibliography{ref}

\begin{thebibliography}{10}

\bibitem{Gyulassy:2004zy}
M.~Gyulassy and L.~McLerran,
\newblock Nucl. Phys. A {\bf 750}, 30 (2005), arXiv:nucl-th/0405013.

\bibitem{Jacobs:2004qv}
P.~Jacobs and X.-N. Wang,
\newblock Prog. Part. Nucl. Phys. {\bf 54}, 443 (2005), arXiv:hep-ph/0405125.

\bibitem{Muller:2012zq}
B.~Muller, J.~Schukraft, and B.~Wyslouch,
\newblock Ann. Rev. Nucl. Part. Sci. {\bf 62}, 361 (2012), arXiv:1202.3233.

\bibitem{Andronic:2015wma}
A.~Andronic {\em et~al.},
\newblock Eur. Phys. J. C {\bf 76}, 107 (2016), arXiv:1506.03981.

\bibitem{Dong:2019byy}
X.~Dong, Y.-J. Lee, and R.~Rapp,
\newblock Ann. Rev. Nucl. Part. Sci. {\bf 69}, 417 (2019), arXiv:1903.07709.

\bibitem{Dong:2019unq}
X.~Dong and V.~Greco,
\newblock Prog. Part. Nucl. Phys. {\bf 104}, 97 (2019).

\bibitem{Cao:2018ews}
S.~Cao {\em et~al.},
\newblock Phys. Rev. C {\bf 99}, 054907 (2019), arXiv:1809.07894.

\bibitem{Rapp:2018qla}
A.~Beraudo {\em et~al.},
\newblock Nucl. Phys. A {\bf 979}, 21 (2018), arXiv:1803.03824.

\bibitem{Xu:2018gux}
Y.~Xu {\em et~al.},
\newblock Phys. Rev. C {\bf 99}, 014902 (2019), arXiv:1809.10734.

\bibitem{Liu:2023rfi}
F.-L. Liu, X.-Y. Wu, S.~Cao, G.-Y. Qin, and X.-N. Wang,
\newblock Phys. Lett. B {\bf 848}, 138355 (2024), arXiv:2304.08787.

\bibitem{Zhao:2023nrz}
J.~Zhao {\em et~al.},
\newblock Phys. Rev. C {\bf 109}, 054912 (2024), arXiv:2311.10621.

\bibitem{Matsui:1986dk}
T.~Matsui and H.~Satz,
\newblock Phys. Lett. B {\bf 178}, 416 (1986).

\bibitem{Satz:2005hx}
H.~Satz,
\newblock J. Phys. G {\bf 32}, R25 (2006), arXiv:hep-ph/0512217.

\bibitem{Vogt:1999cu}
R.~Vogt,
\newblock Phys. Rept. {\bf 310}, 197 (1999).

\bibitem{Wong:2004zr}
C.-Y. Wong,
\newblock Phys. Rev. C {\bf 72}, 034906 (2005), arXiv:hep-ph/0408020.

\bibitem{Rapp:2008tf}
R.~Rapp, D.~Blaschke, and P.~Crochet,
\newblock Prog. Part. Nucl. Phys. {\bf 65}, 209 (2010), arXiv:0807.2470.

\bibitem{CMS:2012gvv}
CMS, S.~Chatrchyan {\em et~al.},
\newblock Phys. Rev. Lett. {\bf 109}, 222301 (2012), arXiv:1208.2826,
\newblock [Erratum: Phys.Rev.Lett. 120, 199903 (2018)].

\bibitem{Andronic:2024oxz}
A.~Andronic {\em et~al.},
\newblock Eur. Phys. J. A {\bf 60}, 88 (2024), arXiv:2402.04366.

\bibitem{Braun-Munzinger:2000csl}
P.~Braun-Munzinger and J.~Stachel,
\newblock Phys. Lett. B {\bf 490}, 196 (2000), arXiv:nucl-th/0007059.

\bibitem{Thews:2000rj}
R.~L. Thews, M.~Schroedter, and J.~Rafelski,
\newblock Phys. Rev. C {\bf 63}, 054905 (2001), arXiv:hep-ph/0007323.

\bibitem{Zhao:2011cv}
X.~Zhao and R.~Rapp,
\newblock Nucl. Phys. A {\bf 859}, 114 (2011), arXiv:1102.2194.

\bibitem{Yao:2018sgn}
X.~Yao and B.~M\"uller,
\newblock Phys. Rev. D {\bf 100}, 014008 (2019), arXiv:1811.09644.

\bibitem{Chen:2018kfo}
B.~Chen,
\newblock Chin. Phys. C {\bf 43}, 124101 (2019), arXiv:1811.11393.

\bibitem{Wu:2024gil}
B.~Wu and R.~Rapp,
\newblock Universe {\bf 10}, 244 (2024), arXiv:2404.09881.

\bibitem{ALICE:2013osk}
ALICE, B.~B. Abelev {\em et~al.},
\newblock Phys. Lett. B {\bf 734}, 314 (2014), arXiv:1311.0214.

\bibitem{PHENIX:2006gsi}
PHENIX, A.~Adare {\em et~al.},
\newblock Phys. Rev. Lett. {\bf 98}, 232301 (2007), arXiv:nucl-ex/0611020.

\bibitem{Du:2017qkv}
X.~Du, M.~He, and R.~Rapp,
\newblock Phys. Rev. C {\bf 96}, 054901 (2017), arXiv:1706.08670.

\bibitem{Liu:2010ej}
Y.~Liu, B.~Chen, N.~Xu, and P.~Zhuang,
\newblock Phys. Lett. B {\bf 697}, 32 (2011), arXiv:1009.2585.

\bibitem{Yao:2020xzw}
X.~Yao, W.~Ke, Y.~Xu, S.~A. Bass, and B.~M\"uller,
\newblock JHEP {\bf 01}, 046 (2021), arXiv:2004.06746.

\bibitem{Eichten:1994gt}
E.~J. Eichten and C.~Quigg,
\newblock Phys. Rev. D {\bf 49}, 5845 (1994), arXiv:hep-ph/9402210.

\bibitem{Ebert:2002pp}
D.~Ebert, R.~N. Faustov, and V.~O. Galkin,
\newblock Phys. Rev. D {\bf 67}, 014027 (2003), arXiv:hep-ph/0210381.

\bibitem{Godfrey:2004ya}
S.~Godfrey,
\newblock Phys. Rev. D {\bf 70}, 054017 (2004), arXiv:hep-ph/0406228.

\bibitem{Eichten:2019gig}
E.~J. Eichten and C.~Quigg,
\newblock Phys. Rev. D {\bf 99}, 054025 (2019), arXiv:1902.09735.

\bibitem{Schroedter:2000ek}
M.~Schroedter, R.~L. Thews, and J.~Rafelski,
\newblock Phys. Rev. C {\bf 62}, 024905 (2000), arXiv:hep-ph/0004041.

\bibitem{Liu:2012tn}
Y.~Liu, C.~Greiner, and A.~Kostyuk,
\newblock Phys. Rev. C {\bf 87}, 014910 (2013), arXiv:1207.2366.

\bibitem{CDF:1998axz}
CDF, F.~Abe {\em et~al.},
\newblock Phys. Rev. D {\bf 58}, 112004 (1998), arXiv:hep-ex/9804014.

\bibitem{CMS:2022sxl}
CMS, A.~Tumasyan {\em et~al.},
\newblock Phys. Rev. Lett. {\bf 128}, 252301 (2022), arXiv:2201.02659.

\bibitem{Chen:2021uar}
B.~Chen, L.~Wen, and Y.~Liu,
\newblock Phys. Lett. B {\bf 834}, 137448 (2022), arXiv:2111.08490.

\bibitem{Zhao:2022auq}
J.~Zhao and P.~Zhuang,
\newblock Eur. Phys. J. A {\bf 61}, 41 (2025), arXiv:2209.13475.

\bibitem{Wu:2023djn}
B.~Wu, Z.~Tang, M.~He, and R.~Rapp,
\newblock Phys. Rev. C {\bf 109}, 014906 (2024), arXiv:2302.11511.

\bibitem{Zhao:2024jkz}
S.~Zhao and M.~He,
\newblock Phys. Lett. B {\bf 861}, 139283 (2025), arXiv:2407.05234.

\bibitem{Cao:2016gvr}
S.~Cao, T.~Luo, G.-Y. Qin, and X.-N. Wang,
\newblock Phys. Rev. C {\bf 94}, 014909 (2016), arXiv:1605.06447.

\bibitem{Luo:2023nsi}
T.~Luo, Y.~He, S.~Cao, and X.-N. Wang,
\newblock Phys. Rev. C {\bf 109}, 034919 (2024), arXiv:2306.13742.

\bibitem{Xing:2024qcr}
W.-J. Xing, S.-Q. Li, S.~Cao, and G.-Y. Qin,
\newblock Phys. Rev. C {\bf 110}, 024903 (2024), arXiv:2404.12601.

\bibitem{Grandchamp:2001pf}
L.~Grandchamp and R.~Rapp,
\newblock Phys. Lett. B {\bf 523}, 60 (2001), arXiv:hep-ph/0103124.

\bibitem{Fries:2008hs}
R.~J. Fries, V.~Greco, and P.~Sorensen,
\newblock Ann. Rev. Nucl. Part. Sci. {\bf 58}, 177 (2008), arXiv:0807.4939.

\bibitem{Greco:2003vf}
V.~Greco, C.~M. Ko, and R.~Rapp,
\newblock Phys. Lett. B {\bf 595}, 202 (2004), arXiv:nucl-th/0312100.

\bibitem{Plumari:2017ntm}
S.~Plumari, V.~Minissale, S.~K. Das, G.~Coci, and V.~Greco,
\newblock Eur. Phys. J. C {\bf 78}, 348 (2018), arXiv:1712.00730.

\bibitem{Cao:2019iqs}
S.~Cao {\em et~al.},
\newblock Phys. Lett. B {\bf 807}, 135561 (2020), arXiv:1911.00456.

\bibitem{Braaten:1993jn}
E.~Braaten, K.-m. Cheung, and T.~C. Yuan,
\newblock Phys. Rev. D {\bf 48}, R5049 (1993), arXiv:hep-ph/9305206.

\bibitem{Berezhnoy:1997fp}
A.~V. Berezhnoy, V.~V. Kiselev, A.~K. Likhoded, and A.~I. Onishchenko,
\newblock Phys. Atom. Nucl. {\bf 60}, 1729 (1997), arXiv:hep-ph/9703341.

\bibitem{Cacciari:2001td}
M.~Cacciari, S.~Frixione, and P.~Nason,
\newblock JHEP {\bf 03}, 006 (2001), arXiv:hep-ph/0102134.

\bibitem{Cacciari:2012ny}
M.~Cacciari {\em et~al.},
\newblock JHEP {\bf 10}, 137 (2012), arXiv:1205.6344.

\bibitem{Cacciari:2015fta}
M.~Cacciari, M.~L. Mangano, and P.~Nason,
\newblock Eur. Phys. J. C {\bf 75}, 610 (2015), arXiv:1507.06197.

\bibitem{Sjostrand:2006za}
T.~Sjostrand, S.~Mrenna, and P.~Z. Skands,
\newblock JHEP {\bf 05}, 026 (2006), arXiv:hep-ph/0603175.

\bibitem{Eskola:2016oht}
K.~J. Eskola, P.~Paakkinen, H.~Paukkunen, and C.~A. Salgado,
\newblock Eur. Phys. J. C {\bf 77}, 163 (2017), arXiv:1612.05741.

\bibitem{Miller:2007ri}
M.~L. Miller, K.~Reygers, S.~J. Sanders, and P.~Steinberg,
\newblock Ann. Rev. Nucl. Part. Sci. {\bf 57}, 205 (2007),
  arXiv:nucl-ex/0701025.

\bibitem{CMS:2017qjw}
CMS, A.~M. Sirunyan {\em et~al.},
\newblock Phys. Lett. B {\bf 782}, 474 (2018), arXiv:1708.04962.

\bibitem{CMS:2017uoy}
CMS, A.~M. Sirunyan {\em et~al.},
\newblock Phys. Rev. Lett. {\bf 119}, 152301 (2017), arXiv:1705.04727.

\bibitem{Chang:1992jb}
C.-H. Chang and Y.-Q. Chen,
\newblock Phys. Rev. D {\bf 48}, 4086 (1993).

\bibitem{Chang:2003cq}
C.-H. Chang, C.~Driouichi, P.~Eerola, and X.~G. Wu,
\newblock Comput. Phys. Commun. {\bf 159}, 192 (2004), arXiv:hep-ph/0309120.

\bibitem{LHCb:2019tea}
LHCb, R.~Aaij {\em et~al.},
\newblock Phys. Rev. D {\bf 100}, 112006 (2019), arXiv:1910.13404.

\bibitem{ParticleDataGroup:2022pth}
Particle Data Group, R.~L. Workman {\em et~al.},
\newblock PTEP {\bf 2022}, 083C01 (2022).

\bibitem{Xing:2021xwc}
W.-J. Xing, G.-Y. Qin, and S.~Cao,
\newblock Phys. Lett. B {\bf 838}, 137733 (2023), arXiv:2112.15062.

\bibitem{Dang:2023tmb}
Y.~Dang, W.-J. Xing, S.~Cao, and G.-Y. Qin,
\newblock Phys. Rev. C {\bf 109}, 064901 (2024), arXiv:2307.14808.

\bibitem{Burnier:2014ssa}
Y.~Burnier, O.~Kaczmarek, and A.~Rothkopf,
\newblock Phys. Rev. Lett. {\bf 114}, 082001 (2015), arXiv:1410.2546.

\bibitem{Combridge:1978kx}
B.~L. Combridge,
\newblock Nucl. Phys. B {\bf 151}, 429 (1979).

\bibitem{Riek:2010fk}
F.~Riek and R.~Rapp,
\newblock Phys. Rev. C {\bf 82}, 035201 (2010), arXiv:1005.0769.

\bibitem{Megias:2007pq}
E.~Megias, E.~Ruiz~Arriola, and L.~L. Salcedo,
\newblock Phys. Rev. D {\bf 75}, 105019 (2007), arXiv:hep-ph/0702055.

\bibitem{Zhang:2003wk}
B.-W. Zhang, E.~Wang, and X.-N. Wang,
\newblock Phys. Rev. Lett. {\bf 93}, 072301 (2004), arXiv:nucl-th/0309040.

\bibitem{Zhang:2018nie}
L.~Zhang, D.-F. Hou, and G.-Y. Qin,
\newblock Phys. Rev. C {\bf 100}, 034907 (2019), arXiv:1812.11048.

\bibitem{Guo:2000nz}
X.-f. Guo and X.-N. Wang,
\newblock Phys. Rev. Lett. {\bf 85}, 3591 (2000), arXiv:hep-ph/0005044.

\bibitem{Majumder:2009ge}
A.~Majumder,
\newblock Phys. Rev. D {\bf 85}, 014023 (2012), arXiv:0912.2987.

\bibitem{Pang:2012he}
L.~Pang, Q.~Wang, and X.-N. Wang,
\newblock Phys. Rev. C {\bf 86}, 024911 (2012), arXiv:1205.5019.

\bibitem{Pang:2018zzo}
L.-G. Pang, H.~Petersen, and X.-N. Wang,
\newblock Phys. Rev. C {\bf 97}, 064918 (2018), arXiv:1802.04449.

\bibitem{Wu:2018cpc}
X.-Y. Wu, L.-G. Pang, G.-Y. Qin, and X.-N. Wang,
\newblock Phys. Rev. C {\bf 98}, 024913 (2018), arXiv:1805.03762.

\bibitem{Wu:2021fjf}
X.-Y. Wu, G.-Y. Qin, L.-G. Pang, and X.-N. Wang,
\newblock Phys. Rev. C {\bf 105}, 034909 (2022), arXiv:2107.04949.

\bibitem{Chang:2015qea}
C.-H. Chang, X.-Y. Wang, and X.-G. Wu,
\newblock Comput. Phys. Commun. {\bf 197}, 335 (2015), arXiv:1507.05176.

\end{thebibliography}

\end{document}